\documentclass[useAMS,usenatbib,twocolumn]{mnras}
\usepackage{amsmath,bm,amssymb}
\usepackage{dcolumn}
\usepackage[utf8]{inputenc}
\usepackage{graphics,graphicx}
\usepackage{float}
\usepackage{threeparttable}
\usepackage{url}
\usepackage{epstopdf}
\usepackage[nameinlink,capitalize]{cleveref}
\usepackage[figurename=Fig.]{caption}
\DeclareCaptionLabelFormat{continued}{#1 #2 - continued}
\usepackage{siunitx}
\usepackage{epstopdf}
\usepackage{booktabs}
\usepackage{longtable}
\usepackage{color}
\usepackage{mathtools}

\title[ExoMol line lists -- {L}: \texorpdfstring{H$_3^+$, H$_2$D$^+$, D$_2$H$^+$ and D$_3^+$.}{H3+, H2D+, D2H+ and D3+.}]{ExoMol line lists -- {L}: High-resolution line lists of \texorpdfstring{H$_3^+$, H$_2$D$^+$, D$_2$H$^+$ and D$_3^+$.}{H3+, H2D+, D2H+ and D3+.}}

\date{\today}

\author[Bowesman et al.]{Charles A. Bowesman$^{1}$, Irina I. Mizus$^{2,3}$, Nikolay F. Zobov$^{3}$, Oleg L. Polyansky$^{1,3}$, \newauthor{J\'{a}nos Sarka$^{4,5}$, Bill Poirier$^{4}$, Marco Pezzella$^{1}$, Sergei N. Yurchenko$^{1}$ and Jonathan Tennyson$^{1}$\thanks{The corresponding author: j.tennyson@ucl.ac.uk}}
\vspace*{4mm}\\
$^1$ Department of Physics and Astronomy, University College London, Gower Street, WC1E 6BT London, UK;\\
$^2$ Holon Institute of Technology, Golomb Street, 52, Holon, 5810201, Israel;\\
$^3$ Institute of Applied Physics, Russian Academy of Science, Ulyanov Street 46, Nizhnii Novgorod, Russia 603950;\\
$^4$ Department of Chemistry and Biochemistry, Texas Tech University, Lubbock, Texas 79409, United States;\\
$^5$ Institute of Chemistry, Eötvös Loránd University, Budapest, Hungary.\\
}

\begin{document}

\label{firstpage}

\maketitle

\pagerange{\pageref{firstpage}--\pageref{lastpage}}

\begin{abstract}
New MiZo line lists are presented for the D$_2$H$^+$ and D$_3^+$ isotopologues of H$_3^+$.
These line lists plus the existing H$_3^+$ MiZATeP and the Sochi H$_2$D$^+$ line lists are updated using empirical energy levels generated using the MARVEL procedure for H$_3^+$, H$_2$D$^+$ and D$_2$H$^+$, and effective Hamiltonian energies for D$_3^+$ for which there is significantly less laboratory data available.
These updates allow accurate frequencies for far infrared lines for these species to be predicted.
Assignments of the energy levels of H$_3^+$ and D$_3^+$ are extended using a combination of high accuracy variational calculations and analysis of transition intensities.
All line lists are made available via www.exomol.com.

\end{abstract}


\section{Introduction}

H$_3^+$ is known to form rapidly in H$_2$ gas following an ionisation event via the strongly exothermic reaction
\begin{equation}
    \centering
    \label{eq:formation}
    \rm{H}_2 + \rm{H}^+_2 \longrightarrow \rm{H}^+_3 + \rm{H}
\end{equation}
which occurs at essentially every collision.
As H$_2$ is common in a variety of astronomical bodies, H$_3^+$ is often the dominant molecular ion.
The first 30 years of H$_3^+$ astronomy has been comprehensively reviewed by \citet{jt800}.
H$_3^+$ formation is stimulated by cosmic ray ionisation of the interstellar medium and by collisions with fast electrons and other charged particles in planetary ionospheres.
H$_3^+$ is also believed to form in the ionosphere of planets through the ionisation of H$_2$ by extreme ultraviolet radiation \citep{16ChGaKo.H3+}.
In gas giant ionospheres, H$_3^+$ acts as a coolant through efficient infrared (IR) emissions \citep{jt489}; indeed it is thought that H$_3^+$ emissions are key to determining the stability limits in hot Jupiter exoplanets \citep{07KoAyMi.H3+}.

The infrared spectrum of H$_3^+$ has been extensively observed in giant planets in our solar system such as Jupiter \citep{jt80,jt142,jt210,17MoOdMe.H3+}, Saturn \citep{93GeJaOk.H3+,08StMiMe.H3+,08StMiLy.H3+}, Uranus \citep{jt127,jt192,jt245,19MeFlSt.H3+} and is believed to be present in Neptune \citep{11MeStMi.H3+,18MeFlSt.H3+} although it is yet to be detected there. 
Its presence can be used as an effective temperature probe here and in other astrophysical settings \citep{22GiFi.H3+}.
H$_3^+$ is similarly expected to be of importance in extrasolar giant planets \citep{16ChGaKo.H3+,15KhShLa.H3+}, such as hot-Jupiters \citep{16LeReSe.H3+}, and an even more prominent feature in the aurorae of brown dwarfs \citep{22GiFi.H3+}; however, it has so far defied observation on these objects.

H$_3^+$ has also been observed in the interstellar medium  (ISM) via absorption in the infrared light of a background star \citep{06Oka.H3+} where it forms through cosmic ray ionisation \citep{96GeOkxx.H3+}.
Hence it is also used in this setting to trace the cosmic ray ionisation rate \citep{12InMcxx.H3+,17HaSiBr.H3+} and similarly primarily relies on IR emissions.
These IR bands lie well within the wavelength range of the instruments onboard the JWST.

In regions where the precursors to H$_3^+$ exist in deuterated forms, namely HD and D$_2$, equivalent reactions occur to that described in \cref{eq:formation} resulting in the formation of the deuterated isotopologues H$_2$D$^+$, D$_2$H$^+$ and D$_3^+$ \citep{22MeHoDe.H3+}.
At low temperatures fractionation drives the preferential formation of isotopically substituted H$_3^+$ \citep{jt344}; indeed, models by \citet{04WaFlde.H3+} suggest that in certain very cold regions D$_3^+$ may be the dominant isotopologue of H$_3^+$!
Spectra of H$_2$D$^+$ \citep{99StVaVa.H3+,03CaVaCe.H3+} and D$_2$H$^+$ \citep{04VaPhYo.H3+} have been observed in interstellar space through pure rotational transitions which lie in the far infrared / THZ region.
However, D$_3^+$ remains undetected, at least in part because its higher symmetry means that, like H$_3^+$, its pure rotational spectrum is very weak.
Elsewhere, electrons provided by H$_3^+$ have been shown to play an important role in the atmospheres of cool white dwarfs \citep{97BeRuLe.H3+}.

Line lists for H$_3^+$ \citep{jt107,jt181} and the associated partition function \citep{jt169,jt337} have played a key role in the astronomical study of this important molecular ion.
In this work we update the MiZATeP H$_3^+$ line list of \citet{jt666} and the older ST1 H$_2$D$^+$ line list of \citet{jt478}.
We do this using updated versions of the \textsc{marvel} (measured active rotation-vibration energy levels) studies due to \citet{13FuSzMa.H3+} and  \citet{13FuSzFa.H3+}.
We present new line lists for D$_2$H$^+$ and D$_3^+$, for which we also use empirical energy levels to improve the accuracy of the transition frequencies between key levels.
These line lists are produced as part of the ExoMol project \citep{jt528}.

\section{Method}

The triatomic discrete variable representation (DVR) nuclear motion code \textsc{DVR3D} \citep{DVR3D} was used previously and here to compute initial energy levels for H$_3^+$ and its deuterated isotopologues as well as Einstein A-coefficients for each transition.
This code, which is based on the use of an exact nuclear motion kinetic energy operator, has been shown to be capable of giving highly accurate results for the H$_3^+$ system \citep{jt236,jt512}.
It is important to note that in the absence of any absolute measurements of transition intensities for the H$_3^+$ system, all models rely on computed values which are thought to be accurate \citep{jt289,jt587}.
It should be noted that \textsc{DVR3D} only provides assignments for the rigorous quantum numbers: $J$, rotationless parity $e/f$ and the interchange symmetry for two identical atoms.
This means that most states in the existing version of the MiZATeP H$_3^+$ \citep{jt666} and  
ST1 H$_2$D$^+$ \citep{jt478} line lists do not have full ro-vibrational labels. We partially address this issue below. 

\textsc{Marvel} \citep{jt412,12FuCsa,jt704} takes assigned, high resolution spectra and uses them to construct empirical energy levels with spectroscopic accuracy and specified uncertainties.
Use of the energy levels can greatly improve the accuracy with which a line list can predict transition frequencies, see \citet{jt828} for a recent example.
For H$_3^+$, H$_2$D$^+$ and D$_2$H$^+$ \textsc{marvel} spectroscopic networks were then constructed using available laboratory spectra in order to obtain empirical energy levels for each species. 
These empirical energy levels were then used to improve our line lists, correcting for obs.--calc. shifts in levels where empirical energy levels are available for comparison.
Such a refinement allows for a subset of the energies to be provided with very high accuracy, as has been demonstrated in similar projects \citep{jt869}.
This allows for high-accuracy transition frequency predictions to be made \citep{jt828}, making the final line list well suited for high-resolution studies \citep{jt835,jt858}.

The \textsc{marvel} process determines an uncertainty for each energy level based on the uncertainties of the input transition that define that level.

\subsection{Spectroscopic Networks}

Transition data for a molecule can be aggregated to construct a spectroscopic network, where the transition frequencies represent the edges of the networks and the energy levels the nodes \citep{jt412,11CsFuxx.marvel}.
The \textsc{marvel} procedure \citep{12FuCsa} achieves this by inverting transition matrices, which yields a set of empirical energy levels with individual uncertainties.
Spectroscopic networks have been constructed in the past for the molecules H$_3^+$ \citep{13FuSzMa.H3+}, H$_2$D$^+$ and D$_2$H$^+$ \citep{13FuSzFa.H3+} but new transition data has since been published.
New transition data allows us to expand the level coverage of the networks and in the case of new data that remeasures existing transitions  to improve the accuracy to which term energies are known.
The recent high-resolution experiments have provided new THz transition data with uncertainties on the order of MHz or kHz, well within the part-per-billion regime.
When all of the transitions that determine a level's term energy are consistent within their experimental uncertainties, the uncertainty on the final term energy will generally be on the same order, but not less than, the smallest uncertainty of the transitions that define the level.
Hence, with the addition of these high-accuracy transition measurements we are able to significantly improve the accuracy of our final energies, in some cases by a few orders of magnitude.

The \textsc{marvel} procedure requires all transitions within a network to be identified by the same set of quantum numbers.
For this purpose, the deuterated isotopologues can be divided into two groups: H$_3^+$ and D$_3^+$ are symmetric-tops belonging to the D$_{\rm 3h}$(M) molecular symmetry group and H$_2$D$^+$ and D$_2$H$^+$ are asymmetric-tops belonging to the C$_{\rm 2v}$(M) group.
These in turn dictate the set of good quantum numbers used to define the levels of the networks.

As symmetric tops, the molecules H$_3^+$ and D$_3^+$ are defined by two primary vibrational modes, symmetric stretching ($\nu_1$) and bending ($\nu_2$).
The bending mode is degenerate however, and as such these species are also described by the vibrational angular momentum quantum number $l_2 = \nu_2, \nu_2 - 2, ..., -\nu_2 + 2, -\nu_2$. 
The energy levels of these species are identified instead by $L_2 = \lvert l_2 \rvert$ in this work.
The rigorous rotational angular momentum quantum number $J$ is used to define the rotational levels of these molecules.
The projection of $J$ along the molecular symmetry axis, $k$ can be used to determine the parity of the energy levels, such that the total parity is the sign of $\left(-1\right)^k$ \citep{13FuSzMa.H3+}.
The quantum number $k$ does not offer a complete description of the system however, due to Coriolis coupling between it and $l_2$.
As such, the value $G = \lvert g \rvert$ is used, where $g = k - l_2$ \citep{62Hougen.H3+}.
This quantum number is important because, as members of the D$_{\rm 3h}$(M) symmetry group, the $A_1^{\prime}$, $A_1^{\prime\prime}$, $A_2^{\prime}$, $A_2^{\prime\prime}$ rovibronic symmetries only exist for these molecules when $G = 3n$, where $n$ is an integer.
In H$_3^+$ the $A_1^{\prime}$ and $A_1^{\prime\prime}$ symmetries do not exist however, as they are determined to have 0 nuclear spin statistical weight \citep{84Watson.H3+}.
This difference arises from the nuclear spin $I$ of constituent atoms, which are $I = \frac{1}{2}$ for hydrogen and $I = 1$ for deuterium.
Consequently H$_3^+$ consists of two nuclear spin isomers while D$_3^+$ has three \citep{87WaFoMc.H3+}; these are shown in \cref{tab:ssw}.

\begin{table}
    \centering
    \caption{The rovibronic symmetries of the isotopologues of H$_3^+$ where $g_{\rm ns}$ is the nuclear spin statistical weight. For H$_3^+$ and D$_3^+$, $\Gamma_{\rm rve}$ are representations of the D$_{\rm 3h}$ point group while for H$_2$D$^+$ and D$_2$H$^+$ they are for C$_{\rm 2v}$.}
    \label{tab:ssw}
    \begin{tabular}{ccccc}
        \hline
        $\Gamma_{\rm rve}$ & Isomer & $g_{\rm ns}$ & Isomer & $g_{\rm ns}$ \\
        \hline
         & \multicolumn{2}{c}{H$_3^+$} & \multicolumn{2}{c}{D$_3^+$} \\
         \hline
        $A_1^{\prime}$, $A_1^{\prime\prime}$ & - & 0 & ortho & 10 \\
        $A_2^{\prime}$, $A_2^{\prime\prime}$ & ortho & 4 & para & 1 \\
        $E^{\prime}$, $E^{\prime\prime}$ & para & 2 & meta & 8 \\
        \hline
         & \multicolumn{2}{c}{H$_2$D$^+$} & \multicolumn{2}{c}{D$_2$H$^+$} \\
        \hline
        $A_1$, $A_2$ & para & 3 & ortho & 12 \\
        $B_1$, $B_2$ & ortho & 9 & para & 6 \\
        \hline
    \end{tabular}
\end{table}

Levels with equivalent vibrational, $J$ and $G$ assignments are differentiated by the $\left(u|l|m\right)$ $U$-notation of \citet{94Watson.H3+}.
$u$ and $l$ are used to identify the upper and lower energy levels of the same assignment, differing in their value of $K = \lvert k \rvert$, and $m$ is used when such a distinction is irrelevant as only one value of $K$ can exist that produces the same $G$.
As such, the energy levels of the species H$_3^+$ and D$_3^+$ are identified by the quantum number set $\left(\nu_1,\nu_2,L_2,J,G,U,K,\Gamma_{\rm rve}\right)$.

As asymmetric-tops, H$_2$D$^+$ and D$_2$H$^+$ are defined by the symmetric stretching, bending and anti-symmetric stretching vibrational quantum numbers $\nu_1$, $\nu_2$ and $\nu_3$.
The quantum number $J$ and its projections onto the $C_2$ axis and the axis perpendicular to the $C_2$ axis in the plane of the molecule, $K_a$ and $K_c$, are the standard ones for labelling the rotational states of an asymmetric top. 

For both asymmetric-tops, the total parity is the sign of $\left(-1\right)^{K_c}$.
A similar expression is used to identify the spin isomers of each asymmetric-top, such that they are labelled ortho when it is positive and para when negative. 
For H$_2$D$^+$ this is expression is $\left(-1\right)^{\nu_3 + K_a}$ and for D$_2$H$^+$ it is $\left(-1\right)^{\nu_3 + K_a + K_c}$.
The ortho and para nuclear spin isomers identify the rovibronic symmetry group of the molecule, $A_1$, $A_2$, $B_1$ or $B_2$, as shown in \cref{tab:ssw}.
Hence, the energy levels of the species H$_2$D$^+$ and D$_2$H$^+$ are identified by the quantum number set $\left(\nu_1,\nu_2,\nu_3,J,K_a,K_c,\Gamma_{\rm rve}\right)$.

Transitions between nuclear spin isomers are forbidden, meaning networks constructed from observational transitions will be split into separate components for each isomer.
Given \textsc{marvel} determines the energies of each level relative to the lowest energy level in the network, which is defined as 0, this presents a problem as any levels of a different nuclear spin isomer to that of the zero-energy level will not have their energies defined.
This is avoided by the introduction of ``magic'' numbers; forbidden transitions are added to the networks to connect the lowest levels of each nuclear spin isomer to each other using calculated energies.
This enables the determination of all energies for connected levels within the networks, relative to the zero-energy level.
For D$_3^+$ this treatment was not needed as the energies were determined through the use of effective Hamiltonian calculations.

\subsubsection{\texorpdfstring{H$_3^+$}{H3+} network}

\citet{13FuSzMa.H3+} conducted a \textsc{marvel} study of H$_3^+$ which combined transition data from 26 sources into a single network.
Since then, five new sets of high-accuracy spectra have been reported  \citep{12BeWoPe.H3+,13HoPeJe.H3+,15PeHoMa.H3+,16JuKoSc.H3+,18GuChLi.H3+,19MaMcxx.H3+} and are detailed below:

\textit{12BeWoPe} \citep{12BeWoPe.H3+}: 3 R-branch transitions in the visible region are provided that had not been included from other sources.

\textit{13HoPeJeSi} \citep{13HoPeJe.H3+}:  10 R-branch transitions in the $\nu_2$ band with MHz and sub-MHz accuracy are reported.
All of these transitions had been previously reported in other sources \citep{81Oka.H3+,98McWaxx.H3+} but are presented here at higher accuracy. 

\textit{15PeHoMaKo} \citep{15PeHoMa.H3+}: A further 10 R-branch transitions in the $\nu_2$ band are reported with MHz and sub-MHz accuracy.
All of the transitions from this source had been observed previously \citep{81Oka.H3+,01LiRaTa.H3+,13WuLiLi.H3+}.

\textit{16JuKoScAs} \citep{16JuKoSc.H3+}: 5 measurements of low-$J$ R-branch transitions in the $\nu_2$ band are provided, all with sub-MHz accuracy.
These transitions had also been observed by \textit{13HoPeJeSi} \citep{13HoPeJe.H3+}.

\textit{18GuChLiPe} \citep{18GuChLi.H3+}: Transition frequencies for 12 R-branch and 4 Q-branch transitions in the $\nu_2$ band are presented.
The majority are reported with sub-MHz accuracy.
All of the transitions in this source had also been observed in other experiments \citep{01LiMcxx.H3+,13HoPeJe.H3+,15PeHoMa.H3+,16JuKoSc.H3+,16PeMcKo.H3+}.

\textit{19MaMc} \citep{19MaMcxx.H3+}: The largest of the new H$_3^+$ data sources, providing 36 measurements of transitions in the $\nu_2 \leftarrow 0$ band, 15 transitions in the $2\nu_2^2 \leftarrow \nu_2$ band and 7 transitions in the $2\nu_2^2 \leftarrow 0$ band.
The data comprises a selection of P-, Q- and R-branch transitions with uncertainties of 4 MHz.
All of the reported transitions had been observed in prior studies \citep{84WaFoMc.H3+,87MaMaMc.H3+,90BaReOk.H3+,90XuGaOk.H3+,94UyGaJa.H3+,98McWaxx.H3+,18GuChLi.H3+,18MaMcSc.H3+}.

In total there were 102 new transitions to be added to the existing set of 1\,610 transitions and a further 7 transitions that had been missed in the original \textsc{marvel} compilation.
The inclusion of the new transitions allowed us to validate 19 previously invalidated transitions.
7 transitions had their assignment corrected based on updated assignments provided in subsequent papers.
A further 9 transitions were reassigned based on apparent mixing identified through comparison with the empirical \textsc{marvel} energies.
2 transition assignments were updated based on suggestions provided by \citet{22Jaquet.H3+}.
\citet{22SaPo} provided a series of assignments for H$_3^+$ which were used to correct the assignments of 41 transitions and to assign 22 previously unassigned transitions.
A further three unassigned transitions from the literature were assigned based on comparison with the \textsc{marvel} energy levels.
The new transition assignments are detailed in \cref{tab:h3p_assignments}.
The transitions updated in this way all involve levels with relatively high energies, in a region where accurate determination of the vibrational state can be challenging.
In total, 59 transitions from the original network were reassigned and an additional 8 had their assignment corrected due to digitisation errors from the original source.

\begin{table*}
    \centering
    \caption{The new transition assignments for transitions from the literature that previously had no assignment or were missing an upper state assignment. Assignments to transitions from \citet{01LiRaTa.H3+} were determined through comparison between the empirical energy levels determined by \textsc{marvel}, while transitions from \citet{09MoGoOk.H3+} were assigned based on assignments produced using the code ScalIT \citep{22SaPo}.}
    \label{tab:h3p_assignments}
    \makebox[1\linewidth][c]{
        \resizebox{1\linewidth}{!}{
            \begin{tabular}{cccccccccccccccccc}
                \hline
                Wavenumber (cm$^{-1}$) & $\nu_1^{\prime}$ & $\nu_2^{\prime}$ & $L_2^{\prime}$ & $J^{\prime}$ & $G^{\prime}$ & $U^{\prime}$ & $K^{\prime}$ & $\Gamma_{\rm rve}^{\prime}$ & $\nu_1^{\prime\prime}$ & $\nu_2^{\prime\prime}$ & $L_2^{\prime\prime}$ & $J^{\prime\prime}$ & $G^{\prime\prime}$ & $U^{\prime\prime}$ & $K^{\prime\prime}$ & $\Gamma_{\rm rve}^{\prime\prime}$ & Source \\
                \hline
                3104.125(10) & 2 & 0 & 0 & 6 & 1 & m & 1 & E$^{\prime\prime}$ & 1 & 0 & 0 & 6 & 2 & m & 2 & E$^{\prime}$ & \citet{01LiRaTa.H3+} \\
                3182.605(5) & 1 & 2 & 2 & 5 & 2 & u & 0 & E$^{\prime}$ & 0 & 2 & 0 & 5 & 1 & m & 1 & E$^{\prime\prime}$ & \citet{01LiRaTa.H3+} \\
                3235.521(5) & 1 & 2 & 2 & 7 & 5 & u & 3 & E$^{\prime\prime}$ & 1 & 1 & 1 & 6 & 5 & l & 6 & E$^{\prime}$ & \citet{01LiRaTa.H3+} \\
                10329.307(10) & 0 & 4 & 2 & 7 & 3 & u & 1 & A$_2^{\prime\prime}$ & 0 & 0 & 0 & 6 & 6 & m & 6 & A$_2^{\prime}$ & \cite{09MoGoOk.H3+} \\
                10462.405(10) & 0 & 5 & 1 & 5 & 6 & m & 5 & A$_2^{\prime\prime}$ & 0 & 0 & 0 & 6 & 6 & m & 6 & A$_2^{\prime}$ & \cite{09MoGoOk.H3+} \\
                10496.287(10) & 2 & 2 & 2 & 6 & 8 & m & 6 & E$^{\prime}$ & 0 & 0 & 0 & 5 & 5 & m & 5 & E$^{\prime\prime}$ & \cite{09MoGoOk.H3+} \\
                10573.997(10) & 0 & 4 & 4 & 5 & 0 & m & 4 & A$_2^{\prime}$ & 0 & 0 & 0 & 4 & 3 & m & 3 & A$_2^{\prime\prime}$ & \cite{09MoGoOk.H3+} \\
                10639.058(10) & 0 & 5 & 1 & 5 & 0 & m & 1 & A$_2^{\prime\prime}$ & 0 & 0 & 0 & 5 & 0 & m & 5 & A$_2^{\prime}$ & \cite{09MoGoOk.H3+} \\
                10666.604(10) & 0 & 5 & 1 & 5 & 3 & l & 4 & A$_2^{\prime}$ & 0 & 0 & 0 & 5 & 3 & m & 3 & A$_2^{\prime\prime}$ & \cite{09MoGoOk.H3+} \\
                10827.764(10) & 0 & 6 & 0 & 3 & 0 & m & 0 & A$_2^{\prime}$ & 0 & 1 & 1 & 2 & 0 & m & 1 & A$_2^{\prime\prime}$ & \cite{09MoGoOk.H3+} \\
                11036.111(10) & 0 & 5 & 1 & 7 & 6 & l & 7 & A$_2^{\prime\prime}$ & 0 & 0 & 0 & 6 & 6 & m & 6 & A$_2^{\prime}$ & \cite{09MoGoOk.H3+} \\
                11046.569(10) & 0 & 5 & 1 & 6 & 5 & l & 6 & E$^{\prime}$ & 0 & 0 & 0 & 5 & 5 & m & 5 & E$^{\prime\prime}$ & \cite{09MoGoOk.H3+} \\
                11048.996(10) & 0 & 5 & 1 & 5 & 4 & l & 5 & E$^{\prime\prime}$ & 0 & 0 & 0 & 4 & 4 & m & 4 & E$^{\prime}$ & \cite{09MoGoOk.H3+} \\
                11114.628(10) & 2 & 2 & 2 & 6 & 3 & u & 1 & A$_2^{\prime\prime}$ & 0 & 0 & 0 & 5 & 0 & m & 5 & A$_2^{\prime}$ & \cite{09MoGoOk.H3+} \\
                11265.189(10) & 1 & 5 & 3 & 3 & 0 & m & 3 & A$_2^{\prime\prime}$ & 0 & 1 & 1 & 2 & 3 & m & 2 & A$_2^{\prime}$ & \cite{09MoGoOk.H3+} \\
                11331.112(10) & 0 & 5 & 5 & 5 & 0 & m & 5 & A$_2^{\prime\prime}$ & 0 & 0 & 0 & 6 & 6 & m & 6 & A$_2^{\prime}$ & \cite{09MoGoOk.H3+} \\
                11556.914(10) & 0 & 5 & 3 & 5 & 3 & m & 0 & A$_2^{\prime}$ & 0 & 0 & 0 & 4 & 3 & m & 3 & A$_2^{\prime\prime}$ & \cite{09MoGoOk.H3+} \\
                11947.074(10) & 2 & 3 & 1 & 5 & 6 & m & 5 & A$_2^{\prime\prime}$ & 0 & 0 & 0 & 6 & 6 & m & 6 & A$_2^{\prime}$ & \cite{09MoGoOk.H3+} \\
                12116.353(10) & 0 & 6 & 0 & 3 & 3 & m & 3 & A$_2^{\prime\prime}$ & 0 & 0 & 0 & 3 & 0 & m & 0 & A$_2^{\prime}$ & \cite{09MoGoOk.H3+} \\
                12331.180(10) & 1 & 4 & 0 & 5 & 0 & m & 0 & A$_2^{\prime}$ & 0 & 0 & 0 & 4 & 3 & m & 3 & A$_2^{\prime\prime}$ & \cite{09MoGoOk.H3+} \\
                12502.614(10) & 0 & 6 & 2 & 2 & 0 & m & 2 & A$_2^{\prime}$ & 0 & 0 & 0 & 3 & 3 & m & 3 & A$_2^{\prime\prime}$ & \cite{09MoGoOk.H3+} \\
                12536.621(10) & 1 & 4 & 2 & 4 & 3 & m & 1 & A$_2^{\prime\prime}$ & 0 & 0 & 0 & 3 & 0 & m & 0 & A$_2^{\prime}$ & \cite{09MoGoOk.H3+} \\
                13056.013(10) & 0 & 6 & 2 & 2 & 3 & m & 1 & A$_2^{\prime\prime}$ & 0 & 0 & 0 & 1 & 0 & m & 0 & A$_2^{\prime}$ & \cite{09MoGoOk.H3+} \\
                13597.367(10) & 0 & 6 & 4 & 1 & 3 & m & 1 & A$_2^{\prime\prime}$ & 0 & 0 & 0 & 1 & 0 & m & 0 & A$_2^{\prime}$ & \cite{09MoGoOk.H3+} \\
                13676.446(10) & 0 & 7 & 1 & 1 & 0 & m & 1 & A$_2^{\prime\prime}$ & 0 & 0 & 0 & 1 & 0 & m & 0 & A$_2^{\prime}$ & \cite{09MoGoOk.H3+} \\
                \hline
            \end{tabular}
        }
    }
\end{table*}

The final set of 1\,719 transitions were passed through \textsc{marvel} for validation, with 1\,656 being validated; these transitions are summarised in \cref{tab:h3p_network_components}.
The network is divided into 39 components, the largest of which comprises 1\,580 transitions which determine 703 empirical energy levels.
Of the remaining components: two consist of 6 transitions; one of 5 transitions; two of 4 transitions; three of 3 transitions; 12 of 2 transitions and the remaining 18 of single transitions.
As these minor components are not connected to the primary component the energies of the levels within them are not determined relative to the zero-energy level and are hence not considered further.

\begin{table}
    \centering    
    \caption{The components of the updated H$_3^+$ \textsc{marvel} network. The network is broken down by the transition data sources and the vibrational bands contained within them, given in the form $\nu_1^{\prime}$, $\nu_2^{\prime}$, $L_2^{\prime}$ - $\nu_1^{\prime\prime}$, $\nu_2^{\prime\prime}$, $L_2^{\prime\prime}$. For each band, the transition energy and total angular momentum $J$ ranges are provided, along with the mean and maximum uncertainties for these transitions. The total number of transitions validated and accessed by \textsc{marvel} (V/A) are also provided.}
    \label{tab:h3p_network_components}
    \makebox[1\linewidth][c]{
        \resizebox{1.1\linewidth}{!}{
            \begin{tabular}{>{\centering\arraybackslash}b{0.14\linewidth}%
                >{\centering\arraybackslash}b{0.12\linewidth}%
                >{\centering\arraybackslash}b{0.12\linewidth}%
                >{\centering\arraybackslash}b{0.23\linewidth}%
                >{\centering\arraybackslash}b{0.345\linewidth}}
                \hline
                Vib. & $J$ range & V/A & Energy range (cm$^{-1}$) & Uncertainty Mean/Max (cm$^{-1}$) \\
                \hline
                \multicolumn{5}{l}{80Oka \citep{80Oka.H3+}} \\
                011 - 000 & 0 - 4 & 14/15 & 2457 - 2918 & 0.013/0.015 \\
                \multicolumn{5}{l}{81Oka \citep{81Oka.H3+}} \\
                011 - 000 & 0 - 6 & 27/30 & 2457 - 3030 & 0.010/0.010 \\
                \multicolumn{5}{l}{84WaFoMcBe \citep{84WaFoMc.H3+}} \\             
                011 - 000 & 0 - 7 & 46/46 & 2217 - 3030 & 0.010/0.010 \\
                \multicolumn{5}{l}{84WaFoMcBe\_CD \citep{84WaFoMc.H3+}} \\
                000 - 000 & 1 - 4 & 5/5 & 173 - 596 & 0.010/0.012 \\
                \multicolumn{5}{l}{87MaMaMcJo \citep{87MaMaMc.H3+}} \\
                011 - 000 & 0 - 10 & 110/113 & 1798 - 3193 & 0.010/0.020 \\
                \multicolumn{5}{l}{89MaFeWaMi \citep{jt83}} \\
                022 - 000 & 1 - 12 & 44/50 & 4540 - 5094 & 0.017/0.041 \\
                \multicolumn{5}{l}{90BaReOk \citep{90BaReOk.H3+}} \\
                011 - 000 & 3 - 9 & 14/14 & 2468 - 2889 & 0.010/0.010 \\
                020 - 011 & 1 - 6 & 14/14 & 2395 - 2685 & 0.010/0.010 \\
                022 - 011 & 0 - 8 & 68/68 & 2090 - 2945 & 0.010/0.040 \\
                111 - 011 & 5 - 6 & 2/2 & 2741 - 2854 & 0.010/0.010 \\
                111 - 100 & 1 - 4 & 21/21 & 2089 - 2771 & 0.010/0.010 \\
                \multicolumn{5}{l}{90NaItSuTa \citep{90NaItSu.H3+}} \\
                011 - 000 & 0 - 3 & 11/12 & 2457 - 2762 & 0.010/0.010 \\
                \multicolumn{5}{l}{90XuGaOk \citep{90XuGaOk.H3+}} \\
                022 - 000 & 1 - 10 & 34/34 & 4557 - 5094 & 0.011/0.040 \\
                \multicolumn{5}{l}{91LeVeCaOk \citep{jt102}} \\
                031 - 000 & 1 - 5 & 3/4 & 6866 - 6878 & 0.012/0.017 \\
                \multicolumn{5}{l}{92XuRoGaOk \citep{92XuRoGa.H3+}} \\
                011 - 000 & 5 - 13 & 30/30 & 2419 - 3291 & 0.010/0.014 \\
                022 - 011 & 3 - 7 & 9/10 & 2893 - 3215 & 0.010/0.010 \\
                022 - 100 & 6 - 7 & 1/1 & 2980 - 2980 & 0.010/0.010 \\
                100 - 000 & 4 - 7 & 9/9 & 2454 - 3282 & 0.010/0.010 \\
                111 - 011 & 0 - 6 & 21/21 & 2437 - 3229 & 0.010/0.010 \\
                120 - 022 & 4 - 5 & 1/1 & 2990 - 2990 & 0.010/0.010 \\
                211 - 111 & 2 - 3 & 1/1 & 3006 - 3006 & 0.010/0.010 \\
                \multicolumn{5}{l}{94MaMcSaWa \citep{94MaMcSa.H3+}} \\
                011 - 000 & 5 - 14 & 67/67 & 1844 - 3643 & 0.014/0.050 \\
                020 - 011 & 1 - 10 & 27/27 & 1934 - 3024 & 0.013/0.050 \\
                022 - 000 & 2 - 8 & 9/11 & 4435 - 4795 & 0.017/0.030 \\
                022 - 011 & 1 - 10 & 57/63 & 1869 - 3293 & 0.013/0.040 \\
                022 - 100 & 5 - 6 & 1/2 & 2836 - 2836 & 0.010/0.010 \\
                031 - 011 & n/a & 0/1 & nan - nan & nan/nan \\
                031 - 020 & 1 - 7 & 2/2 & 2260 - 3002 & 0.010/0.010 \\
                033 - 011 & 3 - 4 & 2/2 & 4553 - 4721 & 0.020/0.030 \\
                033 - 022 & 3 - 7 & 3/4 & 2028 - 3241 & 0.010/0.010 \\
                100 - 000 & 7 - 11 & 4/4 & 3024 - 3277 & 0.010/0.010 \\
                111 - 000 & 6 - 7 & 1/1 & 4949 - 4949 & 0.040/0.040 \\
                111 - 011 & 1 - 9 & 13/18 & 2766 - 3288 & 0.067/0.707 \\
                111 - 100 & 2 - 11 & 36/38 & 1928 - 3204 & 0.011/0.042 \\
                120 - 022 & 6 - 7 & 1/1 & 3206 - 3206 & 0.010/0.010 \\
                122 - 020 & 9 - 10 & 1/1 & 2943 - 2943 & 0.010/0.010 \\
                122 - 022 & n/a & 0/1 & nan - nan & nan/nan \\
                \multicolumn{5}{l}{94UyGaJaOk \citep{94UyGaJa.H3+}} \\
                011 - 000 & 1 - 16 & 73/73 & 2691 - 3579 & 0.011/0.040 \\
                100 - 000 & 8 - 10 & 2/2 & 3412 - 3441 & 0.010/0.010 \\
                \multicolumn{5}{l}{94VeCaGuJo \citep{94VeCaGu.H3+}} \\
                031 - 000 & 1 - 5 & 14/15 & 6807 - 7266 & 0.017/0.040 \\
                \multicolumn{5}{l}{97DiNePoTe \citep{jt193}} \\
                000 - 020 & n/a & 0/1 & nan - nan & nan/nan \\
                011 - 000 & 6 - 14 & 14/16 & 2396 - 3269 & 0.012/0.040 \\
                020 - 011 & 4 - 10 & 9/10 & 2622 - 2998 & 0.013/0.040 \\
                020 - 100 & 9 - 10 & 1/1 & 2893 - 2893 & 0.010/0.010 \\
                022 - 011 & 4 - 13 & 17/24 & 2413 - 3240 & 0.010/0.010 \\
                022 - 100 & 4 - 6 & 2/4 & 2571 - 2709 & 0.010/0.010 \\
                031 - 020 & 1 - 9 & 9/9 & 2458 - 2977 & 0.010/0.010 \\
                031 - 022 & 4 - 6 & 2/2 & 2680 - 2709 & 0.010/0.010 \\
                \hline
            \end{tabular}
        }
    }
\end{table}

\begin{table}
    \ContinuedFloat
    \centering
    \caption{}
    \makebox[1\linewidth][c]{
        \resizebox{1.1\linewidth}{!}{
            \begin{tabular}{>{\centering\arraybackslash}b{0.14\linewidth}%
                >{\centering\arraybackslash}b{0.12\linewidth}%
                >{\centering\arraybackslash}b{0.12\linewidth}%
                >{\centering\arraybackslash}b{0.23\linewidth}%
                >{\centering\arraybackslash}b{0.345\linewidth}}
                \hline
                Vib. & $J$ range & V/A & Energy range (cm$^{-1}$) & Uncertainty Mean/Max (cm$^{-1}$) \\
                \hline\multicolumn{5}{l}{97DiNePoTe \citep{jt193}} \\
                033 - 022 & 0 - 9 & 19/20 & 2484 - 3061 & 0.010/0.010 \\
                100 - 000 & 8 - 8 & 1/1 & 2471 - 2471 & 0.010/0.010 \\
                111 - 011 & 4 - 7 & 5/9 & 2661 - 3203 & 0.020/0.030 \\
                111 - 100 & 4 - 11 & 7/7 & 2458 - 3294 & 0.013/0.030 \\
                120 - 022 & 2 - 6 & 2/2 & 2852 - 3052 & 0.010/0.010 \\
                120 - 111 & 4 - 6 & 2/2 & 2714 - 2817 & 0.025/0.040 \\
                122 - 111 & 2 - 7 & 7/7 & 2403 - 3022 & 0.010/0.010 \\
                211 - 200 & 1 - 6 & 3/3 & 2416 - 2654 & 0.010/0.010 \\
                \multicolumn{5}{l}{00McOk \citep{00McOkx1.H3+}} \\
                120 - 000 & 3 - 4 & 1/1 & 8038 - 8038 & 0.010/0.010 \\
                122 - 000 & 1 - 5 & 27/27 & 7785 - 8163 & 0.014/0.040 \\
                211 - 000 & 4 - 6 & 2/2 & 8053 - 8123 & 0.010/0.010 \\
                \multicolumn{5}{l}{01LiMc \citep{01LiMcxx.H3+}} \\
                011 - 000 & 6 - 7 & 1/1 & 2956 - 2956 & 0.010/0.010 \\
                020 - 011 & 7 - 10 & 3/3 & 2902 - 2977 & 0.017/0.030 \\
                020 - 100 & 8 - 9 & 1/1 & 3003 - 3003 & 0.010/0.010 \\
                022 - 011 & 3 - 8 & 4/4 & 2601 - 2953 & 0.015/0.020 \\
                022 - 100 & 5 - 7 & 3/3 & 2844 - 2980 & 0.010/0.010 \\
                031 - 020 & 3 - 7 & 2/2 & 2596 - 2716 & 0.010/0.010 \\
                033 - 022 & 0 - 5 & 4/4 & 2469 - 2579 & 0.010/0.010 \\
                100 - 000 & 7 - 8 & 1/1 & 2957 - 2957 & 0.030/0.030 \\
                111 - 011 & 5 - 9 & 9/9 & 2437 - 3000 & 0.010/0.010 \\
                111 - 100 & 1 - 9 & 4/4 & 2575 - 2902 & 0.010/0.010 \\
                120 - 022 & 4 - 6 & 2/2 & 2976 - 2990 & 0.010/0.010 \\
                120 - 111 & 5 - 6 & 1/1 & 2817 - 2817 & 0.010/0.010 \\
                122 - 111 & 1 - 6 & 8/8 & 2497 - 2893 & 0.015/0.030 \\
                211 - 111 & 4 - 6 & 2/2 & 2891 - 2903 & 0.010/0.010 \\
                211 - 200 & 5 - 6 & 1/1 & 2654 - 2654 & 0.010/0.010 \\
                \multicolumn{5}{l}{01LiMc\_MAGIC \citep{01LiMcxx.H3+}} \\
                000 - 000 & 0 - 1 & 3/3 & 22.839 - 86.960 & 1.00$\times10^{-9}$/1.00$\times10^{-9}$ \\
                \multicolumn{5}{l}{01LiRaOk \citep{01LiRaTa.H3+}} \\
                011 - 000 & 4 - 16 & 96/96 & 3008 - 3596 & 0.008/0.045 \\
                020 - 011 & 8 - 13 & 6/6 & 2998 - 3283 & 0.007/0.015 \\
                020 - 100 & 9 - 10 & 1/1 & 3111 - 3111 & 0.010/0.010 \\
                022 - 011 & 4 - 12 & 75/75 & 3007 - 3527 & 0.007/0.040 \\
                022 - 100 & 7 - 9 & 4/4 & 3101 - 3269 & 0.013/0.035 \\
                031 - 020 & 5 - 8 & 2/2 & 3003 - 3134 & 0.005/0.005 \\
                033 - 022 & 5 - 6 & 1/1 & 3076 - 3076 & 0.010/0.010 \\
                100 - 000 & 4 - 11 & 22/22 & 3024 - 3575 & 0.009/0.045 \\
                111 - 011 & 0 - 9 & 24/25 & 3009 - 3443 & 0.006/0.015 \\
                111 - 100 & 6 - 10 & 14/14 & 3024 - 3296 & 0.006/0.010 \\
                120 - 020 & 6 - 7 & 1/1 & 3063 - 3063 & 0.015/0.015 \\
                120 - 022 & 5 - 6 & 2/2 & 3018 - 3052 & 0.005/0.005 \\
                122 - 020 & 5 - 5 & 1/1 & 3183 - 3183 & 0.005/0.005 \\
                122 - 111 & 6 - 7 & 2/2 & 3022 - 3236 & 0.005/0.005 \\
                200 - 100 & 6 - 8 & 3/3 & 3077 - 3104 & 0.008/0.010 \\
                \multicolumn{5}{l}{03GoMcOk \citep{03GoMcOk.H3+}} \\
                051 - 000 & 1 - 4 & 7/7 & 11019 - 11576 & 0.011/0.015 \\
                053 - 000 & 1 - 3 & 2/2 & 11044 - 11247 & 0.010/0.010 \\
                055 - 000 & 1 - 3 & 5/5 & 11572 - 11854 & 0.010/0.010 \\
                062 - 000 & 1 - 1 & 1/1 & 12419 - 12419 & 0.010/0.010 \\
                142 - 000 & 1 - 2 & 1/1 & 12246 - 12246 & 0.020/0.020 \\
                231 - 000 & 1 - 3 & 3/3 & 12222 - 12254 & 0.013/0.020 \\
                311 - 000 & 1 - 3 & 3/3 & 11112 - 11504 & 0.010/0.010 \\
                \multicolumn{5}{l}{04MiKrWePl \citep{04MiKrWe.H3+}} \\
                031 - 000 & 1 - 3 & 3/3 & 7193 - 7241 & 0.015/0.026 \\
                \multicolumn{5}{l}{04OkEp \citep{04OkEpxx.H3+}} \\
                011 - 000 & 1 - 7 & 20/20 & 2530 - 3182 & 0.011/0.020 \\
                022 - 000 & 1 - 7 & 10/11 & 4777 - 4987 & 0.015/0.054 \\
                \multicolumn{5}{l}{04OkEp\_CD \citep{04OkEpxx.H3+}} \\
                000 - 000 & 1 - 10 & 68/69 & 7.255 - 1696 & 0.011/0.022 \\
                \multicolumn{5}{l}{08KrBiRePe \citep{08KrBiRe.H3+}} \\
                051 - 000 & 1 - 2 & 2/2 & 11229 - 11244 & 0.010/0.010 \\
                055 - 000 & 0 - 2 & 4/4 & 11594 - 11882 & 0.010/0.010 \\
                062 - 000 & 1 - 2 & 5/5 & 12419 - 13072 & 0.011/0.014 \\
                142 - 000 & 1 - 2 & 2/2 & 12019 - 12087 & 0.010/0.010 \\
                144 - 000 & 1 - 2 & 1/1 & 12898 - 12898 & 0.010/0.010 \\
                231 - 000 & 0 - 1 & 1/1 & 12239 - 12239 & 0.010/0.010 \\
                311 - 000 & 0 - 2 & 4/4 & 11259 - 11511 & 0.010/0.010 \\
                322 - 000 & 1 - 2 & 1/1 & 13333 - 13333 & 0.010/0.010 \\
                \hline
            \end{tabular}
        }
    }
\end{table}

\begin{table}
    \ContinuedFloat
    \centering
    \caption{}
    \makebox[1\linewidth][c]{
        \resizebox{1.1\linewidth}{!}{
            \begin{tabular}{>{\centering\arraybackslash}b{0.14\linewidth}%
                >{\centering\arraybackslash}b{0.12\linewidth}%
                >{\centering\arraybackslash}b{0.12\linewidth}%
                >{\centering\arraybackslash}b{0.23\linewidth}%
                >{\centering\arraybackslash}b{0.345\linewidth}}
                \hline
                Vib. & $J$ range & V/A & Energy range (cm$^{-1}$) & Uncertainty Mean/Max (cm$^{-1}$) \\
                \hline
                \multicolumn{5}{l}{08VeLeAgBe \citep{08VeLeAg.H3+}} \\
                051 - 000 & 1 - 3 & 4/4 & 10730 - 10790 & 0.010/0.010 \\
                062 - 000 & 1 - 3 & 2/2 & 12503 - 13056 & 0.010/0.010 \\
                142 - 000 & 3 - 4 & 1/1 & 12658 - 12658 & 0.010/0.010 \\
                144 - 000 & 1 - 2 & 1/1 & 12898 - 12898 & 0.010/0.010 \\
                222 - 000 & 1 - 3 & 4/4 & 10726 - 10779 & 0.010/0.010 \\
                \multicolumn{5}{l}{09MoGoOk \citep{09MoGoOk.H3+}} \\
                042 - 000 & 6 - 7 & 1/1 & 10329 - 10329 & 0.010/0.010 \\
                044 - 000 & 2 - 5 & 4/4 & 10367 - 10574 & 0.010/0.010 \\
                051 - 000 & 0 - 7 & 38/39 & 10462 - 11619 & 0.010/0.015 \\
                053 - 000 & 1 - 5 & 9/9 & 10657 - 11954 & 0.010/0.010 \\
                055 - 000 & 1 - 6 & 19/19 & 11331 - 12321 & 0.010/0.010 \\
                060 - 000 & 3 - 3 & 1/1 & 12116 - 12116 & 0.010/0.010 \\
                060 - 011 & 2 - 3 & 1/1 & 10828 - 10828 & 0.010/0.010 \\
                062 - 000 & 1 - 3 & 4/4 & 12419 - 13056 & 0.014/0.025 \\
                064 - 000 & 1 - 1 & 1/1 & 13597 - 13597 & 0.010/0.010 \\
                071 - 000 & 1 - 1 & 1/1 & 13676 - 13676 & 0.010/0.010 \\
                140 - 000 & 3 - 5 & 3/3 & 11563 - 12331 & 0.010/0.010 \\
                142 - 000 & 1 - 4 & 5/5 & 12102 - 12658 & 0.010/0.010 \\
                144 - 000 & 1 - 2 & 1/1 & 12898 - 12898 & 0.010/0.010 \\
                153 - 011 & 2 - 3 & 1/1 & 11265 - 11265 & 0.010/0.010 \\
                220 - 000 & 4 - 5 & 1/1 & 11114 - 11114 & 0.010/0.010 \\
                222 - 000 & 0 - 6 & 32/33 & 10322 - 11115 & 0.010/0.010 \\
                231 - 000 & 1 - 6 & 7/7 & 11947 - 12525 & 0.011/0.020 \\
                311 - 000 & 1 - 6 & 11/11 & 10875 - 11892 & 0.010/0.010 \\
                \multicolumn{5}{l}{12BeWoPe \citep{12BeWoPe.H3+}} \\
                144 - 000 & 1 - 2 & 1/1 & 12882 - 12882 & 0.005/0.005 \\
                231 - 000 & 1 - 2 & 1/1 & 12589 - 12589 & 0.005/0.005 \\
                233 - 000 & 1 - 2 & 1/1 & 12620 - 12620 & 0.005/0.005 \\
                \multicolumn{5}{l}{12CrHoSiPe \citep{12CrHoSi.H3+}} \\
                011 - 000 & 1 - 2 & 1/1 & 2726 - 2726 & 0.004/0.004 \\
                \multicolumn{5}{l}{12PaAdAlZo \citep{jt512}} \\
                051 - 000 & 0 - 1 & 2/2 & 10799 - 10832 & 0.005/0.005 \\
                062 - 000 & 0 - 1 & 1/1 & 12413 - 12413 & 0.010/0.010 \\
                071 - 000 & 0 - 1 & 1/1 & 13638 - 13638 & 0.005/0.005 \\
                082 - 000 & 0 - 1 & 1/1 & 15059 - 15059 & 0.005/0.005 \\
                162 - 000 & 1 - 2 & 2/2 & 15130 - 15450 & 0.005/0.005 \\
                164 - 000 & 1 - 2 & 1/1 & 15643 - 15643 & 0.005/0.005 \\
                222 - 000 & 1 - 2 & 1/1 & 10752 - 10752 & 0.005/0.005 \\
                231 - 000 & 1 - 2 & 3/3 & 12374 - 12589 & 0.008/0.010 \\
                233 - 000 & 1 - 2 & 1/1 & 12620 - 12620 & 0.005/0.005 \\
                251 - 000 & 1 - 1 & 1/1 & 15717 - 15717 & 0.005/0.005 \\
                342 - 000 & 1 - 2 & 1/1 & 16506 - 16506 & 0.005/0.005 \\
                \multicolumn{5}{l}{13HoPeJeSi \citep{13HoPeJe.H3+}} \\
                011 - 000 & 1 - 5 & 10/10 & 2691 - 2894 & 2.40$\times10^{-5}$/4.51$\times10^{-5}$ \\
                \multicolumn{5}{l}{13WuLiLiSh \citep{13WuLiLi.H3+}} \\
                011 - 000 & 1 - 4 & 12/12 & 2518 - 2918 & 3.14$\times10^{-4}$/1.00$\times10^{-3}$ \\
                \multicolumn{5}{l}{15PeHoMaKo \citep{15PeHoMa.H3+}} \\
                011 - 000 & 3 - 7 & 10/10 & 2895 - 3030 & 6.93$\times10^{-5}$/1.30$\times10^{-4}$ \\
                \multicolumn{5}{l}{16JuKoScAs \citep{16JuKoSc.H3+}} \\
                011 - 000 & 1 - 3 & 5/5 & 2691 - 2823 & 1.08$\times10^{-5}$/2.20$\times10^{-5}$ \\
                \multicolumn{5}{l}{18GuChLiPe \citep{18GuChLi.H3+}} \\
                011 - 000 & 1 - 6 & 16/16 & 2530 - 3015 & 5.20$\times10^{-5}$/3.00$\times10^{-4}$ \\
                \multicolumn{5}{l}{19MaMc \citep{19MaMcxx.H3+}} \\
                011 - 000 & 0 - 6 & 36/36 & 2217 - 3008 & 1.67$\times10^{-4}$/6.33$\times10^{-4}$ \\
                022 - 000 & 1 - 4 & 7/7 & 4968 - 5094 & 2.16$\times10^{-4}$/6.45$\times10^{-4}$ \\
                022 - 011 & 0 - 5 & 15/15 & 2474 - 2945 & 1.48$\times10^{-4}$/2.16$\times10^{-4}$ \\
                \hline
            \end{tabular}
        }
    }
\end{table}

\subsubsection{\texorpdfstring{H$_2$D$^+$}{H2D+} network}

A \textsc{marvel} network for H$_2$D$^+$ was originally produced by \citet{13FuSzFa.H3+} and contained transition data from 13 sources.
Since then, two new sets of transition measurements have been published by \citet{16JuKoSc.H3+} and \citet{17JuTpMu.H3+}.

\textit{16JuKoScAs} \citep{16JuKoSc.H3+}: Transitions frequencies for 11 $\nu_1$ band transitions are reported with MHz and sub-MHz accuracy.
This source provides significantly higher accuracy measurements of transitions previously observed by \citet{85Amano.H3+}.

\textit{17JuToMuGh} \citep{17JuTpMu.H3+}: 3 pure-rotation transition measurements are provided with sub-MHz or kHz accuracy.

These two new sources and a ``magic'' forbidden transition to connect the nuclear spin isomers, derived from effective Hamiltonian calculations using molecular constants from \citet{06Amano.H3+}, brought the total number of transitions for H$_2$D$^+$ to 210; these are summarised in \cref{tab:h2dp_network_components}.
208 transitions were validated successfully by \textsc{marvel}, yielding a final network comprising 7 components, the largest of which contained 200 transitions that determine 109 unique energy levels.
For the remaining 6 components, one contained 3 transitions while the rest are single transition components; these disconnected components are not considered further here.

\begin{table}
    \centering    
    \caption{The components of the updated H$_2$D$^+$ \textsc{marvel} network. The network is broken down by the transition data sources and the vibrational bands contained within them, given in the form $\nu_1^{\prime}$, $\nu_2^{\prime}$, $\nu_3^{\prime}$ - $\nu_1^{\prime\prime}$, $\nu_2^{\prime\prime}$, $\nu_3^{\prime\prime}$. For each band, the transition energy and total angular momentum $J$ ranges are provided, along with the mean and maximum uncertainties for these transitions. The total number of transitions validated and accessed by \textsc{marvel} (V/A) are also provided.}
    \label{tab:h2dp_network_components}
    \makebox[1\linewidth][c]{
        \resizebox{1.1\linewidth}{!}{
            \begin{tabular}{>{\centering\arraybackslash}b{0.14\linewidth}%
                >{\centering\arraybackslash}b{0.12\linewidth}%
                >{\centering\arraybackslash}b{0.11\linewidth}%
                >{\centering\arraybackslash}b{0.23\linewidth}%
                >{\centering\arraybackslash}b{0.345\linewidth}}
                \hline
                Vib. & $J$ range & V/A & Energy range (cm$^{-1}$) & Uncertainty Mean/Max (cm$^{-1}$) \\
                \hline
                \multicolumn{5}{l}{84AmWa \citep{84AmWaxx.H3+}} \\
                100 - 000 & 0 - 5 & 26/27 & 2841 - 3179 & 0.002/0.004 \\
                \multicolumn{5}{l}{84BoDeDeDe \citep{84BoDeDe.H3+}} \\
                000 - 000 & 1 - 1 & 1/1 & 12.423 - 12.423 & 6.67$\times10^{-6}$/6.67$\times10^{-6}$ \\
                \multicolumn{5}{l}{84WaCoPeWo \citep{84WaCoPe.H3+}} \\
                000 - 000 & 1 - 1 & 1/1 & 12.423 - 12.423 & 3.34$\times10^{-6}$/3.34$\times10^{-6}$ \\
                \multicolumn{5}{l}{85Amano \citep{85Amano.H3+}} \\
                100 - 000 & 0 - 5 & 37/37 & 2839 - 3208 & 0.006/0.040 \\
                \multicolumn{5}{l}{85SaKaHi \citep{85SaKaHi.H3+}} \\
                000 - 000 & 2 - 2 & 1/1 & 5.2032 - 5.2032 & 1.23$\times10^{-6}$/1.23$\times10^{-6}$ \\
                \multicolumn{5}{l}{86FoMcPeWa \citep{86FoMcPe.H3+}} \\
                001 - 000 & 0 - 5 & 42/42 & 2109 - 2602 & 0.006/0.042 \\
                010 - 000 & 0 - 7 & 31/31 & 1838 - 2537 & 0.004/0.005 \\
                \multicolumn{5}{l}{02FaDaKo \citep{jt289}} \\
                002 - 000 & 1 - 2 & 1/1 & 4538 - 4538 & 0.002/0.002 \\
                011 - 000 & 0 - 2 & 3/3 & 4423 - 4513 & 8.67$\times10^{-4}$/0.001 \\
                020 - 000 & 1 - 2 & 4/4 & 4271 - 4394 & 0.001/0.002 \\
                \multicolumn{5}{l}{05AmHi \citep{05AmHixx.H3+}} \\
                000 - 000 & 1 - 3 & 2/3 & 12.423 - 21.563 & 1.31$\times10^{-6}$/1.67$\times10^{-6}$ \\
                \multicolumn{5}{l}{06Amano \citep{06Amano.H3+}} \\
                000 - 000 & 1 - 3 & 3/3 & 85.9513 - 115 & 2.72$\times10^{-5}$/3.34$\times10^{-5}$ \\
                \multicolumn{5}{l}{06Amano\_MAGIC \citep{06Amano.H3+}} \\
                000 - 000 & 0 - 1 & 1/1 & 60.030 - 60.030 & 1.00$\times10^{-5}$/1.00$\times10^{-5}$ \\
                \multicolumn{5}{l}{06HlKoPlKo \citep{06HlKoPl.H3+}} \\
                021 - 000 & 0 - 2 & 3/3 & 6459 - 6491 & 0.021/0.031 \\
                \multicolumn{5}{l}{07AsHuMuKu \citep{jt413}} \\
                021 - 000 & 0 - 3 & 9/9 & 6341 - 6589 & 0.002/0.002 \\
                030 - 000 & 0 - 1 & 4/4 & 6242 - 6331 & 0.002/0.002 \\
                100 - 000 & 0 - 2 & 5/5 & 2947 - 3164 & 0.004/0.005 \\
                120 - 000 & 0 - 2 & 7/7 & 6946 - 7106 & 0.002/0.002 \\
                \multicolumn{5}{l}{08AsRiMuWi \citep{08AsRiMu.H3+}} \\
                000 - 000 & 0 - 1 & 1/1 & 45.701 - 45.701 & 6.67$\times10^{-7}$/6.67$\times10^{-7}$ \\
                \multicolumn{5}{l}{09YoMaMoTa \citep{09YoMaMo.H3+}} \\
                000 - 000 & 0 - 3 & 8/8 & 5.2032 - 115 & 3.27$\times10^{-5}$/2.01$\times10^{-4}$ \\
                \multicolumn{5}{l}{16JuKoScAs \citep{16JuKoSc.H3+}} \\
                100 - 000 & 0 - 2 & 11/11 & 2887 - 3095 & 2.66$\times10^{-5}$/1.50$\times10^{-4}$ \\
                \multicolumn{5}{l}{16JuKoScAs\_CD \citep{16JuKoSc.H3+}} \\
                000 - 000 & 0 - 2 & 4/4 & 66.4066 - 132 & 6.75$\times10^{-5}$/1.51$\times10^{-4}$ \\
                \multicolumn{5}{l}{17JuToMuGh \citep{17JuTpMu.H3+}} \\
                000 - 000 & 0 - 2 & 3/3 & 12.423 - 45.701 & 1.19$\times10^{-6}$/3.07$\times10^{-6}$ \\
                \hline
            \end{tabular}
        }
    }
\end{table}

\subsubsection{\texorpdfstring{D$_2$H$^+$}{D2H+} network}

The D$_2$H$^+$ \textsc{marvel} network was originally published by \citet{13FuSzFa.H3+} and was constructed using transition data from 9 sources.
Four new sets of transition data have subsequently been published \citep{16JuKoSc.H3+,17JuTpMu.H3+,17YuPeAm.H3+,19MaKoMc.H3+} and have now been added to the existing network.

\textit{16JuKoScAs} \citep{16JuKoSc.H3+}: This source provides 10 transitions frequencies observed with sub-MHz accuracy.
8 of these transitions had been previously observed by \citet{84LuAmxx.H3+}.

\textit{17JuToMuGh} \citep{17JuTpMu.H3+}: 3 ground state pure-rotation transition measurements are provided with kHz accuracy.

\textit{17YuPeAmMa} \citep{17YuPeAm.H3+}: This source reports 5 ground state pure-rotation transitions with sub-MHz accuracy.
4 of these transitions had not been reported in other sources, with the other also being observed by \citet{17JuTpMu.H3+}.

\textit{19MaKoMc} \citep{19MaKoMc.H3+}: Transition frequencies for 37 $\nu_1$ band transitions are provided with MHz accuracy.
10 of the transitions in this source are new and had not been reported elsewhere, while the rest had been observed by \citet{84LuAmxx.H3+} and \citet{16JuKoSc.H3+}.

With the addition of 55 new transition measurements, the new network contains 210 transitions.
Two existing transitions that had digitisation errors in their transitions frequencies were updated to the correct values.
A ``magic'' forbidden transition is included to connect the otherwise distinct ortho and para components of the network, using an frequency calculated by \citet{17YuPeAm.H3+}.
The final set of D$_2$H$^+$ transition data is summarised in \cref{tab:d2hp_network_components}.
All of the transitions in the network were validated by \textsc{marvel}, yielding a primary network component containing 200 transitions which define 115 unique energy levels.
The rest of the transitions are present in 7 disconnected components, one of which contains 3 transitions, another 2 transitions and the remainder are single transition components.

\begin{table}
    \centering    
    \caption{The components of the updated D$_2$H$^+$ \textsc{marvel} network. The network is broken down by the transition data sources and the vibrational bands contained within them, given in the form $\nu_1^{\prime}$, $\nu_2^{\prime}$, $\nu_3^{\prime}$ - $\nu_1^{\prime\prime}$, $\nu_2^{\prime\prime}$, $\nu_3^{\prime\prime}$. For each band, the transition energy and total angular momentum $J$ ranges are provided, along with the mean and maximum uncertainties for these transitions. The total number of transitions validated and accessed by \textsc{marvel} (V/A) are also provided.}
    \label{tab:d2hp_network_components}
    \makebox[1\linewidth][c]{
        \resizebox{1.1\linewidth}{!}{
            \begin{tabular}{>{\centering\arraybackslash}b{0.14\linewidth}%
                >{\centering\arraybackslash}b{0.12\linewidth}%
                >{\centering\arraybackslash}b{0.11\linewidth}%
                >{\centering\arraybackslash}b{0.23\linewidth}%
                >{\centering\arraybackslash}b{0.345\linewidth}}
                \hline
                Vib. & $J$ range & V/A & Energy range (cm$^{-1}$) & Uncertainty Mean/Max (cm$^{-1}$) \\
                \hline
                \multicolumn{5}{l}{84LuAm \citep{84LuAmxx.H3+}} \\
                100 - 000 & 0 - 6 & 34/34 & 2638 - 2990 & 0.005/0.005 \\
                \multicolumn{5}{l}{86FoMcWa \citep{86FoMcWa.H3+}} \\
                001 - 000 & 0 - 5 & 35/35 & 1916 - 2291 & 0.005/0.005 \\
                010 - 000 & 0 - 6 & 53/53 & 1782 - 2280 & 0.004/0.005 \\
                \multicolumn{5}{l}{90PoMc \citep{90PoMcxx.H3+}} \\
                010 - 000 & 3 - 4 & 1/1 & 2276 - 2276 & 0.005/0.005 \\
                \multicolumn{5}{l}{02FaDaKoPo \citep{jt289}} \\
                002 - 000 & 0 - 2 & 6/6 & 3994 - 4066 & 6.17$\times10^{-4}$/9.00$\times10^{-4}$ \\
                011 - 000 & 0 - 2 & 6/6 & 3983 - 4122 & 0.002/0.004 \\
                020 - 000 & 0 - 2 & 4/4 & 3847 - 3887 & 0.003/0.004 \\
                \multicolumn{5}{l}{03HiAm \citep{03HiAmxx.H3+}} \\
                000 - 000 & 1 - 1 & 1/1 & 23.071 - 23.071 & 6.00$\times10^{-7}$/6.00$\times10^{-7}$ \\
                \multicolumn{5}{l}{05AmHi \citep{05AmHixx.H3+}} \\
                000 - 000 & 0 - 2 & 3/3 & 23.071 - 49.254 & 1.00$\times10^{-5}$/1.00$\times10^{-5}$ \\
                \multicolumn{5}{l}{06HlPlBa \citep{jt388}} \\
                102 - 000 & 0 - 3 & 3/3 & 6534 - 6536 & 0.005/0.005 \\
                \multicolumn{5}{l}{07AsHuMuKu \citep{jt413}} \\
                102 - 000 & 0 - 2 & 5/5 & 6467 - 6536 & 0.005/0.005 \\
                111 - 000 & 0 - 1 & 1/1 & 6581 - 6581 & 0.005/0.005 \\
                120 - 000 & 0 - 1 & 1/1 & 6482 - 6482 & 0.005/0.005 \\
                \multicolumn{5}{l}{08AsRiMuWi \citep{08AsRiMu.H3+}} \\
                000 - 000 & 0 - 1 & 1/1 & 49.254 - 49.254 & 5.00$\times10^{-7}$/5.00$\times10^{-7}$ \\
                \multicolumn{5}{l}{16JuKoScAs \citep{16JuKoSc.H3+}} \\
                100 - 000 & 0 - 2 & 10/10 & 2684 - 2866 & 8.80$\times10^{-6}$/2.60$\times10^{-5}$ \\
                \multicolumn{5}{l}{17JuToMuGh \citep{17JuTpMu.H3+}} \\
                000 - 000 & 0 - 2 & 3/3 & 23.071 - 49.254 & 2.08$\times10^{-7}$/3.34$\times10^{-7}$ \\
                \multicolumn{5}{l}{17YuPeAmMa \citep{17YuPeAm.H3+}} \\
                000 - 000 & 1 - 3 & 5/5 & 34.646 - 98.311 & 3.34$\times10^{-6}$/3.34$\times10^{-6}$ \\
                \multicolumn{5}{l}{17YuPeAmMa\_MAGIC \citep{17YuPeAm.H3+}} \\
                000 - 000 & 0 - 1 & 1/1 & 34.918 - 34.918 & 1.00$\times10^{-5}$/1.00$\times10^{-5}$ \\
                \multicolumn{5}{l}{19MaKoMc \citep{19MaKoMc.H3+}} \\
                100 - 000 & 0 - 5 & 37/37 & 2588 - 2928 & 6.83$\times10^{-5}$/1.15$\times10^{-4}$ \\
                \hline
            \end{tabular}
        }
    }
\end{table}

\subsection{Effective Hamiltonians}

There have been significantly fewer observations of D$_3^+$ spectra than of the three other isotopologues considered here.
As such, there was insufficient data to build a well-connected network.
In lieu of this, we used effective Hamiltonian constants from \citet{87WaFoMc.H3+} and \citet{94AmChCi.H3+} to calculate energies for the states in the range $J = 0 - 15$, up to a maximum energy of 2676.387 cm$^{-1}$.
These calculations were performed using the program \textsc{pgopher} \citep{PGOPHER} which also provides full state assignments for the levels it computes.
Hence full quantum number assignments were determined via this method for 282 levels within the bands for which constants were available: 188 levels in the 000 band ($\nu_1$, $\nu_2$, $L_2$); 2 levels in the 010 band; 105 in the 011 band; 22 in the 100 band.
The majority of the levels assigned through this method have $J < 10$.
Further assignments were done manually for states at higher energies with the aid of the vibrational band origins published by \citet{94AmChCi.H3+} and the assigned hot and overtone bands published by \citet{95AlWoHi.H3+}.
A further 1045 states were assigned this way and a breakdown of the assigned bands is given in \cref{tab:d3p_assignments}.
\begin{table}
    \centering
    \caption{The assigned vibrational bands of the D$_3^+$ states file, detailing the $J$ range and maximum energy of each band.}
    \label{tab:d3p_assignments}
    \begin{tabular}{cccccc}
        \hline
        $\nu_1$ & $\nu_2$ & $L_2$ & Count & Energy range (cm$^{-1}$) & $J$ range \\
        \hline
        0 & 0 & 0 & 256 & 0 - 4753 & 0 - 15 \\
        0 & 1 & 1 & 380 & 1835 - 6300 & 0 - 15 \\
        0 & 2 & 0 & 99 & 3531 - 6830 & 0 - 15 \\
        0 & 2 & 2 & 253 & 3646 - 7551 & 0 - 15 \\
        0 & 3 & 1 & 11 & 5214 - 5513 & 0 - 3 \\
        0 & 3 & 3 & 6 & 5401 - 5513 & 0 - 1 \\
        0 & 4 & 0 & 1 & 6774 & 0 \\
        0 & 4 & 2 & 2 & 6860 - 6860 & 0 \\
        0 & 4 & 4 & 2 & 7170 - 7170 & 0 \\
        0 & 5 & 1 & 2 & 8298 - 8298 & 0 \\
        0 & 5 & 3 & 2 & 8376 - 8600 & 0 \\
        0 & 5 & 5 & 2 & 8791 - 8791 & 0 \\
        0 & 6 & 0 & 1 & 9696 & 0 \\
        0 & 6 & 2 & 2 & 9731 - 9731 & 0 \\
        1 & 0 & 0 & 140 & 2301 - 5436 & 0 - 15 \\
        1 & 1 & 1 & 99 & 4060 - 7800 & 0 - 14 \\
        1 & 2 & 0 & 4 & 5712 - 5756 & 0 - 1 \\
        1 & 2 & 2 & 4 & 5796 - 6169 & 0 - 4 \\
        1 & 3 & 1 & 2 & 7368 - 7368 & 0 \\
        1 & 3 & 3 & 2 & 7454 - 7536 & 0 \\
        1 & 4 & 0 & 1 & 8864 & 0 \\
        1 & 4 & 2 & 2 & 9031 - 9031 & 0 \\
        1 & 4 & 4 & 2 & 9163 - 9163 & 0 \\
        2 & 0 & 0 & 36 & 4555 - 5178 & 0 - 5 \\
        2 & 1 & 1 & 2 & 6237 - 6237 & 0 \\
        2 & 2 & 0 & 1 & 7831 & 0 \\
        2 & 2 & 2 & 2 & 7894 - 7894 & 0 \\
        2 & 3 & 1 & 2 & 9424 - 9424 & 0 \\
        2 & 3 & 3 & 2 & 9463 - 9555 & 0 \\
        3 & 0 & 0 & 1 & 6761 & 0 \\
        3 & 1 & 1 & 2 & 8366 - 8366 & 0 \\
        3 & 2 & 0 & 1 & 9896 & 0 \\
        3 & 2 & 2 & 2 & 9943 - 9943 & 0 \\
        4 & 0 & 0 & 1 & 8921 & 0 \\
        \hline
    \end{tabular}
\end{table}
These assignments were added to the calculated states file, given that \textsc{DVR3D} only provides assignments for the rigorous quantum numbers: $J$, rotationless parity $e/f$ and interchange of two of the D atoms.

\section{Line list calculations}

\subsection{Updated \texorpdfstring{H$_3^+$}{H3+} and \texorpdfstring{H$_2$D$^+$}{H2D+}}

While the main line lists were computed using \textsc{DVR3D}, we took advantage of calculations by \citet{21SaDaPo.H3+,22SaPo} performed using the variational nuclear motion programs \textsc{ScalIT} \citep{06ChPo,06ChPo2,10ChPo,10ChPo2,14PePo}, and \textsc{GENIUSH} \citep{09MaCzCs,11FaMaCs} to give full quantum number designations.
The approach applied is detailed by \citet{22SaPo} but we provide a quick summary of the main steps here.
First, using Jacobi coordinates \textsc{ScalIT} calculations were carried out in the four blocks of the $G_4$ permutation-inversion (PI) symmetry group.
Due to the very high convergence accuracy, the full G$_{12}$ PI group labels ($\Gamma_{\rm rve}$) were easily assigned unambiguously using the $\Gamma(G_{12}/D_{3h}) \downarrow G_4$ correlation table.
Next, calculations were repeated with the code \textsc{GENIUSH} with slightly lower accuracy, but still sufficient to match the energy levels with the \textsc{ScalIT} ones.
After the vibrational states were labelled $\left(\nu_1,\nu_2,L_2\right)$, vibrational parent labels were semi-automatically assigned to rovibrational states using the rigid rotor decomposition scheme (RRD) implemented in \textsc{GENIUSH} \citep{10MaFaSzCz}.
The RRD overlaps of \textsc{GENIUSH} also help to assign the rotational quantum numbers $\left(J,G,U,K\right)$, as the symmetric top rigid rotor functions are labelled by $K$.
Using this method we were able to label by the quantum number set $\left(\nu_1,\nu_2,L_2,J,G,U,K,\Gamma_{\rm rve}\right)$ an additional 1\,525 states in the MiZATeP H$_3^+$ line list.

The energies of the existing H$_3^+$ and H$_2$D$^+$ line lists were updated with empirical energies where they were known.
For levels with matching assignments in both the states files and corresponding \textsc{marvel} network, the term energies and their uncertainties were set to the values determined by the \textsc{marvel} procedure.
For the levels that did not exist within the molecules' \textsc{marvel} network, the calculated term energies were retained and estimates for their uncertainties were calculated 
as follows
\begin{equation}
    \centering
    \label{eq:calc_unc_estimate}
    \Delta\tilde{E} =
    \begin{dcases}
        0.1, \qquad\text{if } \tilde{E} < 2000 \\
        \lfloor \tilde{E}/2000 \rfloor / 10, \qquad\text{otherwise.} \\
    \end{dcases}
\end{equation}

\subsection{New line lists}

New line lists were computed for D$_2$H$^+$ and D$_3^+$ molecular ions.
Both calculations used the highly accurate global \emph{ab initio} PES, together with adiabatic and relativistic correction surfaces, computed by \citet{jt512,jt526} which were used for the MiZATeP line list.
To calculate transitions intensities the high-accuracy DMS obtained for the H$_3^+$ system was used \citep{jt587}.
This DMS was obtained by fitting $7$ parameters to a polynomial form written in terms of effective charges, see \citet{94RoKuJa.H3+}.
This DMS was centred on the centre-of-mass ensuring the correct treatment of the centre-of-charge -- centre-of-mass displacement which leads to D$_2$H$^+$ (and H$_2$D$^+$) having a permanent dipole moment. 
 
The \textsc{DVR3D} program suite \citep{DVR3D} was used to compute the final line lists.
As part of this project a new module was added to this suite which converts \textsc{DVR3D} results to ExoMol format \citep{jt548} comprising a states and trans file, and allowing spectra to be easily generated using \textsc{ExoCross} \citep{jt708}.
The module reads the energy levels and Einstein coefficients from \textsc{DVR3D}, requiring as input from the user the molecular symmetry, C$_s$ or C$_{2v}$, and the nuclear statistic weight for each symmetry.
The program, called \textsc{LINELIST}, is available as part of the \textsc{DVR3D} program suite from the ExoMol GitHub pages.

As with H$_3^+$ and H$_2$D$^+$, the calculated term energies of D$_2$H$^+$ were updated with empirical values from the new \textsc{marvel} network where available.
Likewise, the unchanged calculated energies of the new D$_2$H$^+$ and D$_3^+$ line lists had uncertainties estimated using \cref{eq:calc_unc_estimate}. 

\subsection{\texorpdfstring{D$_2$H$^+$}{D2H+} nuclear motion calculations}\label{nuc_calc_d2hp}

\textsc{DVR3D} nuclear motion calculations were performed in the following way: with $31$, $31$, and $50$ grid points for two radial and an angular scattering coordinates.
The calculations used Morse-like oscillators with parameters $3.1$, $0.1$, and $0.006$ for $r_1$ (D -- D) radial coordinate, and spherical oscillators with parameters $0$, $0$, and $0.016$ for the $r_2$ (H -- D$_2$ scattering coordinate).
The dimension of the final vibrational Hamiltonians were set to $5000$.
Calculations included all levels up to $15000\,\text{cm}^{-1}$ for $J \leq 25$.
1500 vibrational basis functions calculated using \textsc{DVR3DRJZ} were passed to \textsc{ROTLEV3} for the rotational step of the calculation.
Following \citet{jt236}, nuclear masses $m_{\rm H}=1.007276$ Da and $m_{\rm D}=2.013553$ Da were used for rotational motion and effective masses intermediate between nuclear and atomic masses $m_{\rm H}=1.007537$ Da and $m_{\rm D}=2.013814$ Da were used for the vibrational motion.
This formulation has been shown to allow for non-adiabatic effects in the calculation.
The resulting MiZo line list is complete up to a temperature of $2000$ K.

\subsection{\texorpdfstring{D$_3^+$}{D3+} nuclear motion calculations}\label{nuc_calc_d3p}

\textsc{DVR3D} calculations for D$_3^+$ were performed using the same size grids and Hamiltonians as those specified for D$_2$H$^+$ above ($31$, $31$, and $50$ grid points for two radial and an angular scattering coordinates, correspondingly, and the final Hamiltonian dimensions equal to $5000$).
The calculations used Morse-like oscillators with parameters $2.6$, $0.1$, and $0.006$ for the $r_1$ coordinate, and spherical oscillators with parameters $0$, $0$, and $0.016$ for the r$_2$ coordinate.
Nuclear masses equal to $m_{\rm D}=2.013553$ Da were used for all calculations.

On the basis of nuclear motion calculation results and the DMS by \citet{jt587} a linelist in the frequency region up to $15\,000\,\text{cm}^{-1}$ for a temperature of $800$ K and using the corresponding partition function value $118.7$ together with a list of corresponding energy levels for D$_3^+$ was computed.
This linelist contains transitions between energy states with $J$ values $0 \text{---} 15$ and energies $0 \text{---} 15\,500\,\text{cm}^{-1}$.
The new MiZo D$_3^+$ line list is complete up to a temperature of 800 K.

Statistics for all of the line lists presented here are given in \cref{tab:line_list_stats}.

\begin{table*}
    \centering
    \caption{Statistics for the new line lists, detailing: the maximum $J$ value for each; the maximum temperature up to which the line list is complete; the maximum lower state energy involved in a transition; the maximum upper state energy involved in a transition; the total number of transitions and the total number of transitions between \textsc{marvel}ised states.}
    \label{tab:line_list_stats}
    \begin{tabular}{ccccccc}
        \hline
        Species & J$_{\rm max}$ & T$_{\rm max}$ (K) & E$_{\rm low}$ (cm$^{-1}$) & E$_{\rm high}$ (cm$^{-1}$) & No. Trans & No. \textsc{marvel}ised Trans \\
        \hline
        H$_3^+$ & 37 & 5000 & 25189.70934 & 42000.74357 & 127542657 & 17147 \\
        H$_2$D$^+$ & 20 & 1750 & 6988.91426 & 18496.21669 & 22164810 & 895 \\
        D$_2$H$^+$ & 25 & 2000 & 8253.939824 & 34838.528613 & 2290235000 & 905 \\
        D$_3^+$ & 15 & 800 & 2639.965597 & 17234.642274 & 36078183 & 0 \\
        \hline
    \end{tabular}
\end{table*}

\subsection{States files}

New states files are provided for each of the four species considered here and are formatted using the standard outlined by \citet{jt548}.
H$_3^+$ and D$_3^+$ are identified using the same set of quantum numbers and hence their states files contain the same columns.
Excerpts from the H$_3^+$ and D$_3^+$ states files are provided in \cref{tab:h3pd3p__states}.
Similarly, the states files for H$_2$D$^+$ and D$_2$H$^+$ contain the same set of quantum number columns and an excerpts for them are shown in \cref{tab:h2dp_d2hp_states}.

\begin{table*}
    \centering
    \caption{Excerpts from the H$_3^+$ and D$_3^+$ states files, using the format defined by \citet{jt548}. Note that the zero-energy level of H$_3^+$ with A$_1^{\prime}$ symmetry does not exist and hence has ``nan'' entries for its lifetime and isomer. Any row for which the $\nu_1$, $\nu_2$, $L_2$, $G$, $U$ or $K$ assignment is not known will likewise have ``nan'' entries in these columns.}
    \label{tab:h3pd3p__states}
    {\tt
    \begin{tabular}{ccccccccccccccccc}
        \hline
        $i$ & $\tilde{E}$ & $g_{\rm tot}$ & $J$ & {\rm unc} & $\tau$ & {\rm e/f} & $\Gamma_{\rm rve}$ & {\rm No.} & {\rm Isomer} & $\nu_1$ & $\nu_2$ & $L_2$ & $G$ & $U$ & $K$ & {\rm Source Tag} \\
        \hline
        \multicolumn{17}{c}{{\rm H}$_3^+$} \\
        \hline
        1 & 0.000000 & 0 & 0 & 0.000000 & nan & e & A1' & 1 & nan & 0 & 0 & 0 & 0 & m & 0 & Ma \\
        2 & 2521.410484 & 2 & 0 & 0.000133 & 8.4474e-03 & e & E' & 1 & p & 0 & 1 & 1 & 1 & m & 0 & Ma \\
        3 & 4998.052947 & 2 & 0 & 0.010004 & 2.4871e-03 & e & E' & 2 & p & 0 & 2 & 2 & 2 & m & 0 & Ma \\
        4 & 5554.061000 & 2 & 0 & 0.010000 & 7.3144e-03 & e & E' & 3 & p & 1 & 1 & 1 & 1 & m & 0 & Ca \\
        5 & 7005.974780 & 2 & 0 & 0.010000 & 3.1660e-03 & e & E' & 4 & p & 0 & 3 & 1 & 1 & m & 0 & Ca \\
        6 & 7870.229810 & 2 & 0 & 0.010000 & 2.5047e-03 & e & E' & 5 & p & 1 & 2 & 2 & 2 & m & 0 & Ca \\
        7 & 8488.013160 & 2 & 0 & 0.010000 & 6.6336e-03 & e & E' & 6 & p & 2 & 1 & 1 & 1 & m & 0 & Ca \\
        8 & 9113.041390 & 2 & 0 & 0.010000 & 2.0836e-03 & e & E' & 7 & p & 0 & 4 & 2 & 2 & m & 0 & Ca \\
        9 & 9653.699850 & 2 & 0 & 0.010000 & 1.1122e-03 & e & E' & 8 & p & 1 & 3 & 1 & 1 & m & 0 & Ca \\
        10 & 9997.183130 & 2 & 0 & 0.010000 & 1.0631e-03 & e & E' & 9 & p & 0 & 4 & 4 & 4 & m & 0 & Ca \\
        \hline
        \multicolumn{17}{c}{{\rm D}$_3^+$} \\
        \hline
        1 & 0.0000000 & 10 & 0 & 0.0000000 & inf & + & A1' & 1 & o & 0 & 0 & 0 & 0 & m & 0 & EH \\
        2 & 2301.1985530 & 10 & 0 & 0.1000000 & 9.6693e+00 & + & A1' & 2 & o & 1 & 0 & 0 & 0 & m & 0 & Ca \\
        3 & 3530.6654530 & 10 & 0 & 0.1000000 & 2.4171e-02 & + & A1' & 3 & o & 0 & 2 & 0 & 0 & m & 0 & Ca \\
        4 & 4554.7949320 & 10 & 0 & 0.1000000 & 3.2217e+00 & + & A1' & 4 & o & 2 & 0 & 0 & 0 & m & 0 & Ca \\
        5 & 5400.9592930 & 10 & 0 & 0.1000000 & 6.7375e-03 & + & A1' & 5 & o & 0 & 3 & 3 & 3 & m & 0 & Ca \\
        6 & 5711.9821150 & 10 & 0 & 0.1000000 & 1.4546e-02 & + & A1' & 6 & o & 1 & 2 & 0 & 0 & m & 0 & Ca \\
        7 & 6761.1836300 & 10 & 0 & 0.1000000 & 1.3900e+00 & + & A1' & 7 & o & 3 & 0 & 0 & 0 & m & 0 & Ca \\
        8 & 6774.1319250 & 10 & 0 & 0.1000000 & 8.2880e-03 & + & A1' & 8 & o & 0 & 4 & 0 & 0 & m & 0 & Ca \\
        9 & 7453.9598120 & 10 & 0 & 0.1000000 & 7.3275e-03 & + & A1' & 9 & o & 1 & 3 & 3 & 3 & m & 0 & Ca \\
        10 & 8375.5588740 & 10 & 0 & 0.1000000 & 6.3728e-03 & + & A1' & 10 & o & 0 & 5 & 3 & 3 & m & 0 & Ca \\
        \hline
    \end{tabular}
    }
    \begin{tablenotes}
        \item $i$: State counting number;
        \item $\tilde{E}$: Term value (in cm$^{-1}$);
        \item $g_{\rm tot}$: Total state degeneracy;
        \item $J$: Total angular momentum quantum number;
        \item unc: Estimated uncertainty of energy level (in cm$^{-1}$);
        \item $\tau$: Radiative lifetime (in seconds);
        \item e/f: Rotationless parity;
        \item $\Gamma_{\rm rve}$: $D_{\rm 3h}$ symmetry group;
        \item No.: Symmetry group block counting number;
        \item Isomer: Nuclear spin isomer; 
        \item $\nu_1$: Symmetric stretching quantum number;
        \item $\nu_2$: Bending quantum number;
        \item $L_2$: Vibrational angular momentum quantum number;
        \item $G$: Absolute value of $k - l_2$;
        \item $U$: $U$-notation of \citet{94Watson.H3+};
        \item $K$: Rotational angular momentum quantum number;
        \item Source Tag: The method used to generate the term value; ``Ma'' for \textsc{marvel}ised energies, ``EH'' for energies from effective Hamiltonian calculations and ``Ca'' for energies calculated using \textsc{DVR3D}.
    \end{tablenotes}
\end{table*}

\begin{table*}
    \centering
    \caption{Excerpts from the H$_2$D$^+$ and D$_2$H$^+$ states files, using the format defined by \citet{jt548}. Any row for which the $\nu_1$, $\nu_2$, $\nu_3$, $K_a$ or $K_c$ assignment is not known will have ``nan'' entries in these columns.}
    \label{tab:h2dp_d2hp_states}
    {\tt 
    \begin{tabular}{ccccccccccccccccc}
        \hline
        $i$ & $\tilde{E}$ & $g_{\rm tot}$ & $J$ & {\rm unc} & $\tau$ & {\rm +/-} & $\Gamma_{\rm rve}$ & {\rm No.} & {\rm Isomer} & $\nu_1$ & $\nu_2$ & $\nu_3$ & $K_a$ & $K_c$ & {\rm Source Tag} \\
        \hline
        \multicolumn{16}{c}{{\rm H}$_2$D$^+$} \\
        \hline
        1 & 0.0000000 & 3 & 0 & 0.0000000 & inf & + & A1 & 1 & p & 0 & 0 & 0 & 0 & 0 & Ma \\
        2 & 2205.8771127 & 3 & 0 & 0.0050000 & 5.7143e-02 & + & A1 & 2 & p & 0 & 1 & 0 & 0 & 0 & Ma \\
        3 & 2992.5022357 & 3 & 0 & 0.0000050 & 1.8706e-02 & + & A1 & 3 & p & 1 & 0 & 0 & 0 & 0 & Ma \\
        4 & 4287.4732000 & 3 & 0 & 0.2000000 & 1.8812e-02 & + & A1 & 4 & p & 0 & 2 & 0 & 0 & 0 & Ca \\
        5 & 4602.6196400 & 3 & 0 & 0.2000000 & 3.1868e-03 & + & A1 & 5 & p & 0 & 0 & 2 & 0 & 0 & Ca \\
        6 & 5039.7666100 & 3 & 0 & 0.2000000 & 1.9520e-02 & + & A1 & 6 & p & 1 & 1 & 0 & 0 & 0 & Ca \\
        7 & 5877.2574100 & 3 & 0 & 0.2000000 & 1.0870e-02 & + & A1 & 7 & p & 2 & 0 & 0 & 0 & 0 & Ca \\
        8 & 6287.6671127 & 3 & 0 & 0.0020000 & 6.4627e-03 & + & A1 & 8 & p & 0 & 3 & 0 & 0 & 0 & Ma \\
        9 & 6645.7008700 & 3 & 0 & 0.3000000 & 2.5443e-03 & + & A1 & 9 & p & 0 & 1 & 2 & 0 & 0 & Ca \\
        10 & 6991.5781127 & 3 & 0 & 0.0020000 & 9.4938e-03 & + & A1 & 10 & p & 1 & 2 & 0 & 0 & 0 & Ma \\
        \hline
        \multicolumn{16}{c}{{\rm D}$_2$H$^+$} \\
        \hline
        1 & 0.0000000 & 12 & 0 & 0.0000000 & inf & + & A1 & 1 & o & 0 & 0 & 0 & 0 & 0 & Ma \\
        2 & 1968.1622648 & 12 & 0 & 0.0050000 & 1.9317e-02 & + & A1 & 2 & o & 0 & 1 & 0 & 0 & 0 & Ma \\
        3 & 2736.9754969 & 12 & 0 & 0.0000040 & 1.1633e-02 & + & A1 & 3 & o & 1 & 0 & 0 & 0 & 0 & Ma \\
        4 & 3821.2081670 & 12 & 0 & 0.1000000 & 1.2227e-02 & + & A1 & 4 & o & nan & nan & nan & nan & nan & Ca \\
        5 & 4042.7721648 & 12 & 0 & 0.0009000 & 1.2305e-02 & + & A1 & 5 & o & 0 & 0 & 2 & 0 & 0 & Ma \\
        6 & 4648.7587400 & 12 & 0 & 0.2000000 & 6.4298e-03 & + & A1 & 6 & o & nan & nan & nan & nan & nan & Ca \\
        7 & 5385.3522270 & 12 & 0 & 0.2000000 & 6.0840e-03 & + & A1 & 7 & o & nan & nan & nan & nan & nan & Ca \\
        8 & 5579.1928220 & 12 & 0 & 0.2000000 & 8.0766e-03 & + & A1 & 8 & o & nan & nan & nan & nan & nan & Ca \\
        9 & 6008.5163100 & 12 & 0 & 0.3000000 & 4.8418e-03 & + & A1 & 9 & o & nan & nan & nan & nan & nan & Ca \\
        10 & 6432.5596700 & 12 & 0 & 0.3000000 & 6.6492e-03 & + & A1 & 10 & o & nan & nan & nan & nan & nan & Ca \\
        \hline
    \end{tabular}
    }
    \begin{tablenotes}
        \item $i$: State counting number;
        \item $\tilde{E}$: Term value (in cm$^{-1}$);
        \item $g_{\rm tot}$: Total state degeneracy;
        \item $J$: Total angular momentum quantum number;
        \item unc: Estimated uncertainty of energy level (in cm$^{-1}$);
        \item $\tau$: Radiative lifetime (in seconds);
        \item +/-: Total parity;
        \item $\Gamma_{\rm rve}$: $C_{\rm 2v}$ symmetry group;
        \item No.: Symmetry group block counting number;
        \item Isomer: Nuclear spin isomer; 
        \item $\nu_1$: Symmetric stretching quantum number;
        \item $\nu_2$: Bending quantum number;
        \item $\nu_3$: Asymmetrical stretching quantum number;
        \item $K_a$: Rotational angular momentum quantum number;
        \item $K_c$: Rotational angular momentum quantum number;
        \item Source Tag: The method used to generate the term value; ``Ma'' for \textsc{marvel}ised energies, ``EH'' for energies from effective Hamiltonian calculations and ``Ca'' for energies calculated using \textsc{DVR3D}.
    \end{tablenotes}
\end{table*}

All entries in the new states files are marked with a source tag, indicating the method that was used to determine the final term energy for that level.
Several values for the source tag can occur in ExoMol line lists \citep{jt835} but only three are in use here: ``Ma'' for \textsc{marvel}ised energies, ``EH'' for energies from effective Hamiltonian calculations and ``Ca'' for energies calculated using \textsc{DVR3D}.
For all levels marked ``Ca'' in all four states files an uncertainty estimate was calculated using \cref{eq:calc_unc_estimate}.

All new and updated states files are ordered on increasing $J$, $\Gamma_{\rm rve}$ and energy.
The existing H$_2$D$^+$ states file was already formatted this way, while H$_3^+$ was not and has been changed.
Hence for H$_3^+$ the state counting number assigned to each level has been reassigned and the references to these values in the corresponding trans file have accordingly been updated.
Excerpts from the trans files for the new D$_2$H$^+$ and D$_3^+$ line lists can be seen in \cref{tab:trans_file_excerpts}.

\begin{table}
    \centering
    \caption{Excerpts from the trans files for the new line lists for D$_2$H$^+$ and D$_3^+$.}
    \label{tab:trans_file_excerpts}
    {\tt
    \begin{tabular}{cccccc}
        \hline
        \multicolumn{3}{c}{{\rm D}$_2$H$^+$} & \multicolumn{3}{c}{{\rm D}$_3^+$} \\
        $f$ & $i$ & {\rm A}$_{fi}$ (s$^{-1}$) & $f$ & $i$ & {\rm A}$_{fi}$ (s$^{-1}$) \\
        \hline
        1401 & 1 & 3.30142477E-03 & 226 & 16 & 5.65667085E-09 \\
        1402 & 2 & 2.19499998E-03 & 229 & 16 & 2.93922054E-02 \\
        3 & 1402 & 1.59575873E-01 & 216 & 226 & 1.30736782E-07 \\
        1403 & 2 & 1.80478736E-03 & 1023 & 226 & 4.95756630E-07 \\
        3 & 1403 & 4.59507405E-02 & 1025 & 226 & 5.90452177E-07 \\
        1404 & 2 & 9.63358275E-02 & 1028 & 226 & 6.58440901E-02 \\
        1404 & 3 & 3.99647847E-03 & 2 & 212 & 1.03419866E-01 \\
        1405 & 4 & 2.49561352E-03 & 1013 & 212 & 2.22162053E-02 \\
        5 & 1405 & 2.09336332E-03 & 229 & 216 & 3.39285830E-02 \\
        6 & 1405 & 3.28445005E-01 & 1043 & 216 & 4.03805710E-07 \\
        \hline
    \end{tabular}
    }
    \begin{tablenotes}
        \item $f$: The upper state counting number from the corresponding states file.
        \item $i$: The lower state counting number from the corresponding states file.
        \item A$_{fi}$: The Einstein A coefficient.
    \end{tablenotes}
\end{table}

In general all molecular states should be characterised by three rigorous quantum numbers: $J$, parity and the overall symmetry, which also determines the nuclear spin state (ortho/para/meta).
All calculations have $J$ and parity rigorously determined; due to the full treatment by \textsc{DVR3D} of the C$_{\rm 2v}$ symmetry, the symmetry and nuclear spin statistics are also correctly determined for H$_2$D$^+$ and D$_2$H$^+$.
This is not the case for H$_3^+$ and D$_3^+$ however, as the program does not include a full treatment for a D$_{\rm 3h}$ Hamiltonian.
The symmetry was initially output for all molecular species considered here under C$_{\rm 2v}$ symmetry, but was subsequently mapped to the appropriate D$_{\rm 3h}$ representation for H$_3^+$ and D$_3^+$ for states with quantum number assignments.
Due to the nuclear spin statistical weight of the A$_1^{\prime}$, A$_1^{\prime\prime}$ symmetries in H$_3^+$ being 0, it was possible to map all of the C$_{\rm 2v}$ symmetries to the appropriate D$_{\rm 3h}$ representation \citep{jt666}.
For D$_3^+$, where D$_{\rm 3h}$ symmetry and quantum number assignment was not carried out, the entries retain the C$_{\rm 2v}$ symmetries output by \textsc{DVR3D}.
These C$_{\rm 2v}$ symmetries are given as lower-case in the states file, to avoid confusion between the A$_1$ and A$_2$ irreducible representations under C$_{\rm 2v}$ symmetry (formatted as ``a1'' and ``a2'') and the A$_1^{\prime}$, A$_1^{\prime\prime}$, A$_2^{\prime}$ and A$_2^{\prime\prime}$ irreducible representations under D$_{\rm 3h}$ symmetry.

\subsection{Partition Functions}

Calculating the partition functions of the new line lists requires the determination of the total degeneracy of each state considered, as show in \cref{eq:partition_function}:
\begin{equation}
    \centering
    \label{eq:partition_function}
    z = \sum_{i} g_{{\rm tot}, i} exp\left(-\frac{E_i}{k_BT}\right),
\end{equation}
where $g_{\rm tot}$ is the total degeneracy of the $i$th level, $E_i$ its energy, $k_B$ is the Boltzmann constant and T is the temperature.
The total degeneracy of a level depends directly on the nuclear spin statistical weight, $g_{\rm ns}$:
\begin{equation}
    \centering
    \label{eq:degeneracy}
    g_{\rm tot} = g_{\rm ns}\left(2J + 1\right).
\end{equation}
The nuclear spin statistical weights represent the number of nuclear spin functions that yield each symmetry.
We follow the ExoMol and HITRAN convention of including the full nuclear spin degeneracy in our partition functions.
In the case of H$_3^+$, where each atom consists of one proton with $I = \frac{1}{2}$, Fermi statistics apply which results in no allowed configurations corresponding to the A$_1^{\prime}$ or A$_1^{\prime\prime}$ representations.
Accordingly, the $g_{\rm ns}$ values for these representations of H$_3^+$ are 0, as shown in \cref{tab:ssw}.
For D$_3^+$ however, each Deuterium atom has $I = 1$ and is hence a boson, meaning additional representations can occur that give rise to the ``meta'' nuclear spin isomer.

Our mappings of the C$_{\rm 2v}$ representations output by \textsc{DVR3D} to the full D$_{\rm 3h}$ symmetry are complete for all states presented in the H$_3^+$ line list, but not for the states of D$_3^+$.
Hence, we must also consider the appropriate statistical weights to use when determining the degeneracy of the D$_3^+$ states left with C$_{\rm 2v}$ symmetries.
As the D$_{\rm 3h}$ representations A$_1^{\prime}$, A$_1^{\prime\prime}$, A$_2^{\prime}$, A$_2^{\prime\prime}$, E$^{\prime}$, E$^{\prime\prime}$ correspond to A$_1$, A$_2$, B$_2$, B$_1$, A$_1 \oplus {\rm B}_2$, A$_2 \oplus {\rm B}_1$ respectively under C$_{\rm 2v}$ symmetry, we can calculate population weighted average statistical weights for each C$_{\rm 2v}$ representation.
These are calculated knowing that approximately two thirds of all D$_3^+$ levels will be E$^{\prime}$ and E$^{\prime\prime}$, due to the constraint that A$_1^{\prime}$, A$_1^{\prime\prime}$, A$_2^{\prime}$ and A$_2^{\prime\prime}$ levels only occur when $G = 3n$, where $n$ is an integer; see also \citet{jt118}.
Accordingly, for the A$_1$ and A$_2$ representations in C$_{\rm 2v}$ symmetry, one third will be A$_1^{\prime}$ or A$_1^{\prime\prime}$ under D$_{\rm 3h}$ symmetry while the remaining two thirds will be E$^{\prime}$ and E$^{\prime\prime}$.
The equivalent population ratio is true for B$_1$ and B$_2$ states and the corresponding A$_2^{\prime}$ and A$_2^{\prime\prime}$ representations.
These population ratios are then weighted using the weights from \cref{tab:ssw}, yielding the $g_{\rm ns}$ values shown in \cref{tab:d3p_c2v_ssw}.
These adjusted weights were used to calculate the total degeneracies of the states of the D$_3^+$ line list for which C$_{\rm 2v}$ symmetries are used and are included in the new states file.

\begin{table}
    \centering
    \caption{Adjusted nuclear spin statistical weights for the levels of D$_3^+$ for which C$_{\rm 2v}$ symmetries are used.}
    \label{tab:d3p_c2v_ssw}
    \begin{tabular}{ccc}
        \hline
        $\Gamma_{\rm rve}$ & $g_{\rm ns}$ \\
        \hline
        A$_1$, A$_2$ & 6 \\
        B$_1$, B$_2$ & 3 \\
        \hline
    \end{tabular}
\end{table}

Partition functions are given in \cref{tab:partition_functions} for a series of temperatures up to the value of T$_{\rm max}$ for each line list.
Partition functions for D$_3^+$ had previously been calculated by \citet{jt337} and their values are in close agreement with those presented here.
For H$_2$D$^+$ however, the partition functions differ from those given by \citet{jt478}.
This is due to the different nuclear spin statistical weights used in their earlier work that did not consider the spin of the Deuterium atom.
The degeneracies given in the H$_2$D$^+$ line list have been updated to use the nuclear spin statistical weights quoted in \cref{tab:ssw} and are thus consistent with the partition function computed using \cref{eq:partition_function} and the same convention.

\begin{table}
    \centering
    \caption{The partition functions for each new line list, calculated for a range of temperatures up to each line list's maximum temperature, given in \cref{tab:line_list_stats}.}
    \label{tab:partition_functions}
    \begin{tabular}{ccccc}
        \hline
        Temp. (K) & H$_3^+$ & H$_2$D$^+$ & D$_2$H$^+$ & D$_3^+$ \\
        \hline
        10 & 6.0000$\times10^{-4}$ & 3.0181$\times10^{0}$ & 1.2153$\times10^{1}$ & 1.0235$\times10^{1}$ \\
        50 & 2.0215$\times10^{0}$ & 1.5141$\times10^{1}$ & 3.7355$\times10^{1}$ & 2.7175$\times10^{1}$ \\
        100 & 7.3972$\times10^{0}$ & 4.5354$\times10^{1}$ & 9.7278$\times10^{1}$ & 6.9095$\times10^{1}$ \\
        200 & 2.0733$\times10^{1}$ & 1.2604$\times10^{2}$ & 2.6721$\times10^{2}$ & 1.9085$\times10^{2}$ \\
        300 & 3.7609$\times10^{1}$ & 2.2937$\times10^{2}$ & 4.8756$\times10^{2}$ & 3.4904$\times10^{2}$ \\
        400 & 5.7650$\times10^{1}$ & 3.5231$\times10^{2}$ & 7.5053$\times10^{2}$ & 5.3858$\times10^{2}$ \\
        500 & 8.0581$\times10^{1}$ & 4.9357$\times10^{2}$ & 1.0549$\times10^{3}$ & 7.6022$\times10^{2}$ \\
        600 & 1.0640$\times10^{2}$ & 6.5391$\times10^{2}$ & 1.4048$\times10^{3}$ & 1.0188$\times10^{3}$ \\
        700 & 1.3533$\times10^{2}$ & 8.3572$\times10^{2}$ & 1.8080$\times10^{3}$ & 1.3215$\times10^{3}$ \\
        800 & 1.6783$\times10^{2}$ & 1.0426$\times10^{3}$ & 2.2746$\times10^{3}$ & 1.6761$\times10^{3}$ \\
        900 & 2.0442$\times10^{2}$ & 1.2787$\times10^{3}$ & 2.8162$\times10^{3}$ & \\
        1000 & 2.4578$\times10^{2}$ & 1.5489$\times10^{3}$ & 3.4456$\times10^{3}$ & \\
        1250 & 3.7478$\times10^{2}$ & 2.4087$\times10^{3}$ & 5.4967$\times10^{3}$ & \\
        1500 & 5.5089$\times10^{2}$ & 3.6069$\times10^{3}$ & 8.4351$\times10^{3}$ & \\
        1750 & 7.8938$\times10^{2}$ & 5.2510$\times10^{3}$ & 1.2563$\times10^{4}$ & \\
        2000 & 1.1086$\times10^{3}$ & & 1.8242$\times10^{4}$ & \\
        3000 & 3.7032$\times10^{3}$ & & & \\
        4000 & 1.0239$\times10^{4}$ & & & \\
        5000 & 2.4177$\times10^{4}$ & & & \\
        \hline
    \end{tabular}
\end{table}

\section{Spectra}

\subsection{Rotational Spectra}

Purely rotational transitions within the vibrational ground states are important for detections of H$_3^+$ and particularly its deuterated isotopologues in the interstellar medium.
This is due to the low temperatures in such regions driving the majority of the level populations to low-energy states.
Predicted transition frequencies for such transitions are given in \cref{tab:predicted_trans_symmetric} for H$_3^+$ and D$_3^+$ and in \cref{tab:predicted_trans_asymmetric} for H$_2$D$^+$ and D$_2$H$^+$, based on the term energies derived from the \textsc{marvel} networks presented here.
While the purely rotational transitions of H$_3^+$ are ``forbidden'' due to the lack of a permanent electric dipole moment, it is believed that a small, temporary dipole moment can be induced due to the distortion of the molecule from equilibrium geometry under rotation about the C$_2$ molecular axis \citep{86PaOkxx.H3+,jt72}.

The D$_3^+$ states file contains 43 extremely long-lived states with radiative lifetimes greater than $10^{10}$ s and can be considered meta-stable.
In the extreme case, two states are calculated to have radiative lifetimes greater than $10^{18}$ s, a duration greater than the current age of the universe.
All of these meta-stable states are in the vibrational ground state and have $G = J$ or $G = J-1$.
The same is true for the meta-stable states of H$_3^+$, though only one of their radiative lifetimes is in excess of $10^{10}$ s.
In both species, this can cause molecules to become ``trapped'' in these states in collision free environments with consequences for both laboratory measurements \citep{jt340} and for possible maser action.

\subsection{Mid and Far Infrared Spectra}

\begin{figure}
    \centering    
    \caption{Far infrared spectra for H$_3^+$ and its deuterated isotopologues, computed using the program \textsc{ExoCross} \citep{jt708} at 200 K between 10 - 1000 cm$^{-1}$ (1 mm - 10 $\mu$m).}
    \label{fig:Far_IR}
    \includegraphics[width=\linewidth]{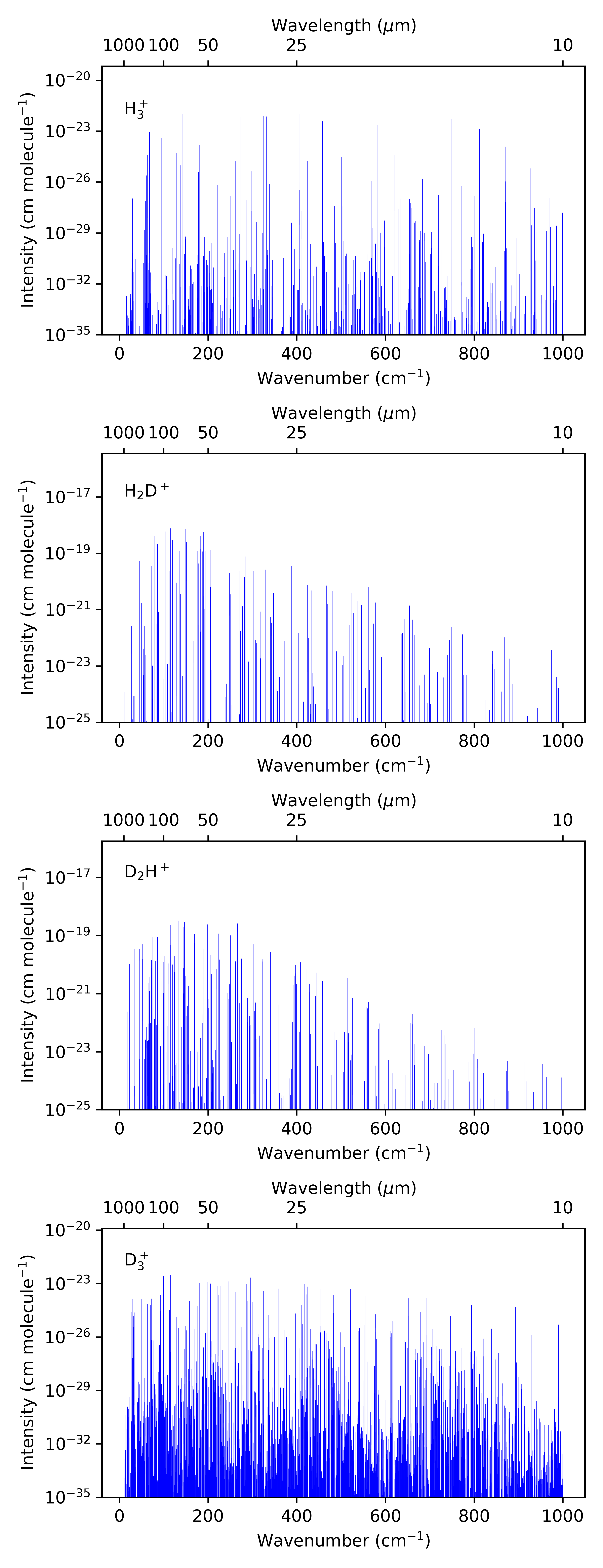}
\end{figure}

\begin{figure}
    \centering    
    \caption{Mid infrared spectra for H$_3^+$ and its deuterated isotopologues, computed using the program \textsc{ExoCross} \citep{jt708} at 200 K between 1000 - 5000 cm$^{-1}$ (10 - 2 $\mu$m).}
    \label{fig:Mid_IR}
    \includegraphics[width=\linewidth]{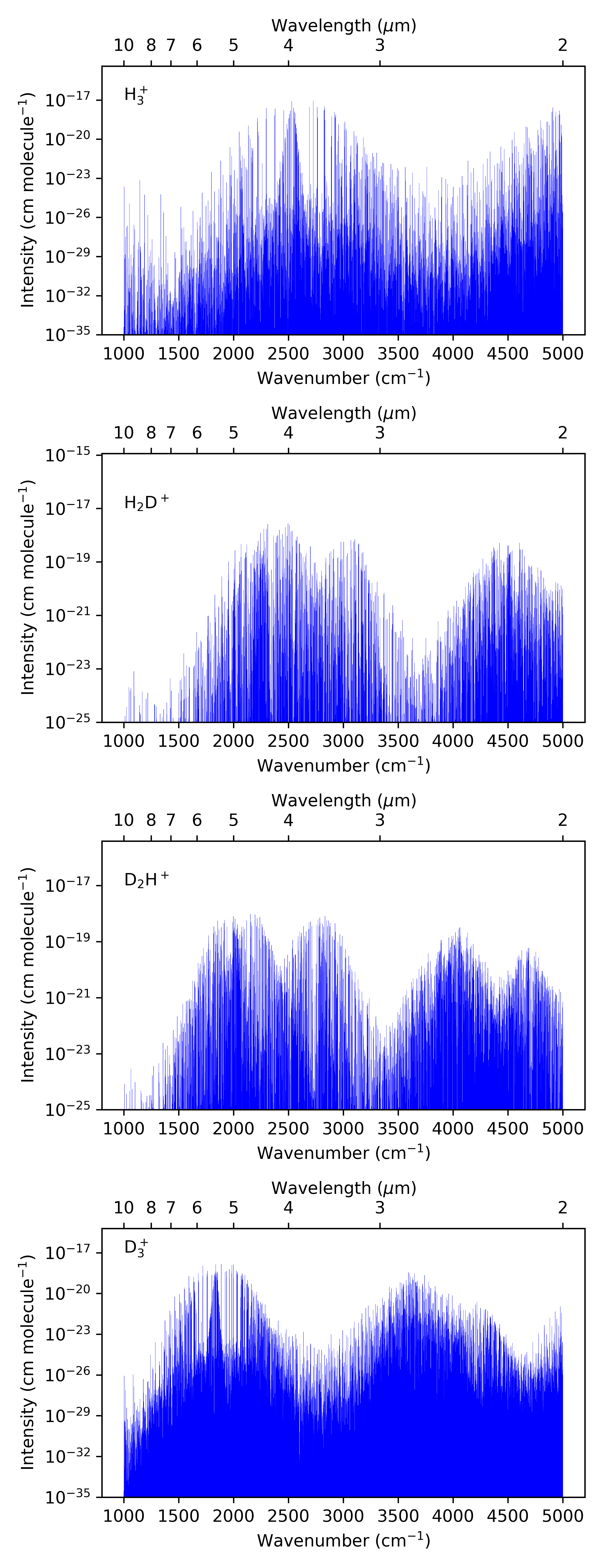}
\end{figure}

\cref{fig:Far_IR} shows example stick spectra for H$_3^+$ and its deuterated isotopologues in the far infrared region from 10 - 1000 cm$^{-1}$.
These spectra were generated using the program \textsc{ExoCross} \citep{jt708} and were computed for a temperature of 200 K.
The predicted transitions frequencies listed in \cref{tab:predicted_trans_symmetric,tab:predicted_trans_asymmetric} that are expected to be of importance in the ISM fall within this range.
Likewise, spectra were computed for the mid infrared region between 1000 - 5000 cm$^{-1}$ for all species considered here and are shown in \cref{fig:Mid_IR}.
These cover several wavelength ranges where H$_3^+$ has been observed in Jupiter's aurorae, such as the K band region centred at 2.2 $\mu$m \citep{14UnKaTa.H3+} and the L band centred on 3.5 $\mu$m \citep{jt108}.
This also covers the wavelength regions observed using the Jupiter infrared Auroral Mapper (JIRAM) instrument onboard NASA's Juno spacecraft \citep{19DiAdMu.H3+,19MiDiMo.H3+} and those covered by the Near infrared Camera (NIRCam) long-wavelength channel onboard JWST.

Comparisons between computed spectra and real observations cannot be made for H$_2$D$^+$, D$_2$H$^+$ and D$_3^+$, as no measured transition intensities have been published.
\cite{98McWaxx.H3+} presented infrared absorption spectra of the $\nu_2$ fundamental band of H$_3^+$ and provided integrated intensities, which were well reproduced by the original line list \citep{jt666}.

\begin{table*}
    \centering
    \caption{Predicted pure-rotation transition wavenumbers and Einstein A coefficients for transitions involving low-energy lower levels for H$_3^+$ and D$_3^+$. For all upper and lower levels included here, $\nu_1 = \nu_2 = L_2 = 0$. Values quoted in brackets for energies are the uncertainty in the last digit.}
    \label{tab:predicted_trans_symmetric}
    \begin{tabular}{rccccccrcccccr}
        \hline
         & & \multicolumn{6}{c}{Upper Level} & \multicolumn{6}{c}{Lower Level} \\
        \multicolumn{1}{c}{Wavenumber (cm$^-1$)} & A$_{fi}$ (s$^{-1}$) & $J$ & $G$ & $U$ & $K$ & $\Gamma_{\rm rve}$ & Term Energy (cm$^-1$) & $J$ & $G$ & $U$ & $K$ & $\Gamma_{\rm rve}$ & Term Energy (cm$^-1$) \\
        \hline
        \multicolumn{14}{c}{H$_3^+$} \\
        \hline
        105.17639(37) & 4.2570$\times10^{-7}$ & 2 & 2 & m & 2 & E$^{\prime}$ & 169.29739(37) & 1 & 1 & m & 1 & E$^{\prime\prime}$ & 64.121000(0) \\
        68.05481(37) & 5.6141$\times10^{-7}$ & 2 & 1 & m & 1 & E$^{\prime\prime}$ & 237.352195(50) & 2 & 2 & m & 2 & E$^{\prime}$ & 169.29739(37) \\
        325.46535(40) & 3.4982$\times10^{-5}$ & 3 & 1 & m & 1 & E$^{\prime\prime}$ & 494.76274(14) & 2 & 2 & m & 2 & E$^{\prime}$ & 169.29739(37) \\
        190.66548(40) & 1.7422$\times10^{-5}$ & 3 & 2 & m & 2 & E$^{\prime}$ & 428.01768(39) & 2 & 1 & m & 1 & E$^{\prime\prime}$ & 237.352195(50) \\
        201.52843(29) & 7.3590$\times10^{-5}$ & 3 & 0 & m & 0 & A$_2^{\prime}$ & 516.88036(23) & 3 & 3 & m & 3 & A$_2^{\prime\prime}$ & 315.35192(18) \\
        66.74506(42) & 2.6423$\times10^{-6}$ & 3 & 1 & m & 1 & E$^{\prime\prime}$ & 494.76274(14) & 3 & 2 & m & 2 & E$^{\prime}$ & 428.01768(39) \\
        405.55623(42) & 3.2067$\times10^{-4}$ & 4 & 1 & m & 1 & E$^{\prime\prime}$ & 833.57391(14) & 3 & 2 & m & 2 & E$^{\prime}$ & 428.01768(39) \\
        7.27431(27) & 2.5602$\times10^{-9}$ & 4 & 4 & m & 4 & E$^{\prime}$ & 502.03705(23) & 3 & 1 & m & 1 & E$^{\prime\prime}$ & 494.76274(14) \\
        273.70656(44) & 1.7953$\times10^{-4}$ & 4 & 2 & m & 2 & E$^{\prime}$ & 768.46930(42) & 3 & 1 & m & 1 & E$^{\prime\prime}$ & 494.76274(14) \\
        \hline
        \multicolumn{14}{c}{D$_3^+$} \\
        \hline
        53.3000(28) & 1.3067$\times10^{-8}$ & 2 & 2 & m & 2 & E$^{\prime}$ & 85.6210(20) & 1 & 1 & m & 1 & E$^{\prime\prime}$ & 32.3210(20) \\\
        33.7470(28) & 1.6552$\times10^{-8}$ & 2 & 1 & m & 1 & E$^{\prime\prime}$ & 119.3680(20) & 2 & 2 & m & 2 & E$^{\prime}$ & 85.6210(20) \\
        163.7280(28) & 1.0739$\times10^{-6}$ & 3 & 1 & m & 1 & E$^{\prime\prime}$ & 249.3490(20) & 2 & 2 & m & 2 & E$^{\prime}$ & 85.6210(20) \\
        96.5440(28) & 5.5135$\times10^{-7}$ & 3 & 2 & m & 2 & E$^{\prime}$ & 215.9120(20) & 2 & 1 & m & 1 & E$^{\prime\prime}$ & 119.3680(20) \\
        29.2810(28) & 3.1041$\times10^{-8}$ & 3 & -3 & m & 3 & A$_1^{\prime\prime}$ & 159.8680(20) & 2 & 0 & m & 0 & A$_1^{\prime}$ & 130.5870(20) \\
        100.5990(28) & 2.1970$\times10^{-6}$ & 3 & 0 & m & 0 & A$_2^{\prime}$ & 260.4660(20) & 3 & 3 & m & 3 & A$_2^{\prime\prime}$ & 159.8670(20) \\
        272.6900(28) & 1.4422$\times10^{-5}$ & 4 & 0 & m & 0 & A$_1^{\prime}$ & 432.5580(20) & 3 & -3 & m & 3 & A$_1^{\prime\prime}$ & 159.8680(20) \\
        33.4370(28) & 8.0083$\times10^{-8}$ & 3 & 1 & m & 1 & E$^{\prime\prime}$ & 249.3490(20) & 3 & 2 & m & 2 & E$^{\prime}$ & 215.9120(20) \\
        205.6600(28) & 1.0260$\times10^{-5}$ & 4 & 1 & m & 1 & E$^{\prime\prime}$ & 421.5720(20) & 3 & 2 & m & 2 & E$^{\prime}$ & 215.9120(20) \\
        5.6660(28) & 3.0205$\times10^{-10}$ & 4 & 4 & m & 4 & E$^{\prime}$ & 255.0150(20) & 3 & 1 & m & 1 & E$^{\prime\prime}$ & 249.3490(20) \\
        139.0820(28) & 5.7366$\times10^{-6}$ & 4 & 2 & m & 2 & E$^{\prime}$ & 388.4310(20) & 3 & 1 & m & 1 & E$^{\prime\prime}$ & 249.3490(20) \\
        \hline
    \end{tabular}
\end{table*}

\begin{table*}
    \centering
    \caption{Predicted pure-rotation transition wavenumbers and Einstein A coefficients for transitions involving low-energy lower levels for H$_2$D$^+$ and D$_2$H$^+$. For all upper and lower levels included here, $\nu_1 = \nu_2 = \nu_3 = 0$. Values quoted in brackets for the transition frequencies are the uncertainty in the last digit.}
    \label{tab:predicted_trans_asymmetric}
    \begin{tabular}{rcccccrccccr}
        \hline
         & & \multicolumn{5}{c}{Upper Level} & \multicolumn{5}{c}{Lower Level} \\
        Wavenumber (cm$^-1$) & A$_{fi}$ (s$^{-1}$) & $J$ & $K_a$ & $K_c$ & $\Gamma_{\rm rve}$ & Term Energy (cm$^-1$) & $J$ & $K_a$ & $K_c$ & $\Gamma_{\rm rve}$ & Term Energy (cm$^-1$) \\
        \hline
        \multicolumn{12}{c}{H$_2$D$^+$} \\
        \hline
        45.7011127(4) & 4.0400$\times10^{-3}$ & 1 & 0 & 1 & A$_2$ & 45.7011127(4) & 0 & 0 & 0 & A$_1$ & 0.0000000(0) \\
        85.951350(33) & 3.0300$\times10^{-2}$ & 2 & 0 & 2 & A$_1$ & 131.652463(33) & 1 & 0 & 1 & A$_2$ & 45.7011127(4) \\
        178.1566(36) & 1.6600$\times10^{-2}$ & 2 & 2 & 0 & A$_1$ & 223.8577(36) & 1 & 0 & 1 & A$_2$ & 45.7011127(4) \\
        12.422639(14) & 1.2200$\times10^{-4}$ & 1 & 1 & 0 & B$_2$ & 72.452486(10) & 1 & 1 & 1 & B$_1$ & 60.029846(10) \\
        78.829277(23) & 1.8800$\times10^{-2}$ & 2 & 1 & 2 & B$_2$ & 138.859124(21) & 1 & 1 & 1 & B$_1$ & 60.029846(10) \\
        103.482549(23) & 4.2400$\times10^{-2}$ & 2 & 1 & 1 & B$_1$ & 175.935034(21) & 1 & 1 & 0 & B$_2$ & 72.452486(10) \\
        87.0021(36) & 3.2700$\times10^{-3}$ & 2 & 2 & 1 & A$_2$ & 218.6545(36) & 2 & 0 & 2 & A$_1$ & 131.652463(33) \\
        119.7622(20) & 8.5300$\times10^{-2}$ & 3 & 0 & 3 & A$_2$ & 251.4146(20) & 2 & 0 & 2 & A$_1$ & 131.652463(33) \\
        244.6884(70) & 4.7600$\times10^{-2}$ & 3 & 2 & 1 & A$_2$ & 376.3409(70) & 2 & 0 & 2 & A$_1$ & 131.652463(33) \\
        37.075911(30) & 1.0700$\times10^{-3}$ & 2 & 1 & 1 & B$_1$ & 175.935034(21) & 2 & 1 & 2 & B$_2$ & 138.859124(21) \\
        115.201469(33) & 7.3200$\times10^{-2}$ & 3 & 1 & 3 & B$_1$ & 254.060593(26) & 2 & 1 & 2 & B$_2$ & 138.859124(21) \\
        319.4843(70) & 2.3600$\times10^{-2}$ & 3 & 3 & 1 & B$_1$ & 458.3434(70) & 2 & 1 & 2 & B$_2$ & 138.859124(21) \\
        150.2252(20) & 1.5900$\times10^{-1}$ & 3 & 1 & 2 & B$_2$ & 326.1602(20) & 2 & 1 & 1 & B$_1$ & 175.935034(21) \\
        283.8945(70) & 3.7500$\times10^{-2}$ & 3 & 3 & 0 & B$_2$ & 459.8296(70) & 2 & 1 & 1 & B$_1$ & 175.935034(21) \\
        5.2032(50) & 1.1300$\times10^{-5}$ & 2 & 2 & 0 & A$_1$ & 223.8577(36) & 2 & 2 & 1 & A$_2$ & 218.6545(36) \\
        136.1238(79) & 7.7800$\times10^{-2}$ & 3 & 2 & 2 & A$_1$ & 354.7783(70) & 2 & 2 & 1 & A$_2$ & 218.6545(36) \\
        27.5569(41) & 9.5600$\times10^{-6}$ & 3 & 0 & 3 & A$_2$ & 251.4146(20) & 2 & 2 & 0 & A$_1$ & 223.8577(36) \\
        152.4832(79) & 1.1200$\times10^{-1}$ & 3 & 2 & 1 & A$_2$ & 376.3409(70) & 2 & 2 & 0 & A$_1$ & 223.8577(36) \\
        103.3637(73) & 8.2700$\times10^{-3}$ & 3 & 2 & 2 & A$_1$ & 354.7783(70) & 3 & 0 & 3 & A$_2$ & 251.4146(20) \\
        151.38(10) & 1.8100$\times10^{-1}$ & 4 & 0 & 4 & A$_1$ & 402.80(10) & 3 & 0 & 3 & A$_2$ & 251.4146(20) \\
        329.972(10) & 6.7000$\times10^{-2}$ & 4 & 2 & 2 & A$_1$ & 581.386(10) & 3 & 0 & 3 & A$_2$ & 251.4146(20) \\
        527.62(10) & 4.6500$\times10^{-3}$ & 4 & 4 & 0 & A$_1$ & 779.04(10) & 3 & 0 & 3 & A$_2$ & 251.4146(20) \\
        72.0997(20) & 4.1900$\times10^{-3}$ & 3 & 1 & 2 & B$_2$ & 326.1602(20) & 3 & 1 & 3 & B$_1$ & 254.060593(26) \\
        149.6189(53) & 1.7400$\times10^{-1}$ & 4 & 1 & 4 & B$_2$ & 403.6795(53) & 3 & 1 & 3 & B$_1$ & 254.060593(26) \\
        205.7690(70) & 3.1700$\times10^{-3}$ & 3 & 3 & 0 & B$_2$ & 459.8296(70) & 3 & 1 & 3 & B$_1$ & 254.060593(26) \\
        391.51(10) & 5.2800$\times10^{-2}$ & 4 & 3 & 2 & B$_2$ & 645.57(10) & 3 & 1 & 3 & B$_1$ & 254.060593(26) \\
        \hline
        \multicolumn{12}{c}{D$_2$H$^+$} \\
        \hline
        49.2542648(2) & 3.3014$\times10^{-3}$ & 1 & 1 & 1 & A$_2$ & 49.2542648(2) & 0 & 0 & 0 & A$_1$ & 0.0000000(0) \\
        23.071309(14) & 5.0885$\times10^{-4}$ & 1 & 1 & 0 & B$_2$ & 57.989629(10) & 1 & 0 & 1 & B$_1$ & 34.918320(10) \\
        75.341734(30) & 1.0592$\times10^{-2}$ & 2 & 1 & 2 & B$_2$ & 110.260054(28) & 1 & 0 & 1 & B$_1$ & 34.918320(10) \\
        52.4637522(33) & 2.0691$\times10^{-3}$ & 2 & 0 & 2 & A$_1$ & 101.7180170(33) & 1 & 1 & 1 & A$_2$ & 49.2542648(2) \\
        132.809827(10) & 4.4605$\times10^{-2}$ & 2 & 2 & 0 & A$_1$ & 182.064091(10) & 1 & 1 & 1 & A$_2$ & 49.2542648(2) \\
        121.174679(68) & 4.4481$\times10^{-2}$ & 2 & 2 & 1 & B$_1$ & 179.164308(67) & 1 & 1 & 0 & B$_2$ & 57.989629(10) \\
        34.6460731(47) & 1.2985$\times10^{-3}$ & 2 & 1 & 1 & A$_2$ & 136.3640901(34) & 2 & 0 & 2 & A$_1$ & 101.7180170(33) \\
        98.3110072(57) & 2.4763$\times10^{-2}$ & 3 & 1 & 3 & A$_2$ & 200.0290242(47) & 2 & 0 & 2 & A$_1$ & 101.7180170(33) \\
        68.904254(73) & 4.5854$\times10^{-3}$ & 2 & 2 & 1 & B$_1$ & 179.164308(67) & 2 & 1 & 2 & B$_2$ & 110.260054(28) \\
        85.840024(73) & 1.4620$\times10^{-2}$ & 3 & 0 & 3 & B$_1$ & 196.100078(67) & 2 & 1 & 2 & B$_2$ & 110.260054(28) \\
        185.781201(73) & 5.4468$\times10^{-2}$ & 3 & 2 & 1 & B$_1$ & 296.041254(67) & 2 & 1 & 2 & B$_2$ & 110.260054(28) \\
        45.700001(11) & 2.2792$\times10^{-3}$ & 2 & 2 & 0 & A$_1$ & 182.064091(10) & 2 & 1 & 1 & A$_2$ & 136.3640901(34) \\
        146.951256(94) & 6.2260$\times10^{-2}$ & 3 & 2 & 2 & A$_1$ & 283.315346(94) & 2 & 1 & 1 & A$_2$ & 136.3640901(34) \\
        72.137151(95) & 1.7139$\times10^{-3}$ & 3 & 1 & 2 & B$_2$ & 251.301459(67) & 2 & 2 & 1 & B$_1$ & 179.164308(67) \\
        17.964933(11) & 5.6565$\times10^{-6}$ & 3 & 1 & 3 & A$_2$ & 200.0290242(47) & 2 & 2 & 0 & A$_1$ & 182.064091(10) \\
        55.201381(95) & 3.5511$\times10^{-3}$ & 3 & 1 & 2 & B$_2$ & 251.301459(67) & 3 & 0 & 3 & B$_1$ & 196.100078(67) \\
        121.1511(50) & 5.0954$\times10^{-2}$ & 4 & 1 & 4 & B$_2$ & 317.2512(50) & 3 & 0 & 3 & B$_1$ & 196.100078(67) \\
        83.286322(94) & 8.5394$\times10^{-3}$ & 3 & 2 & 2 & A$_1$ & 283.315346(94) & 3 & 1 & 3 & A$_2$ & 200.0290242(47) \\
        115.707967(94) & 4.3474$\times10^{-2}$ & 4 & 0 & 4 & A$_1$ & 315.736992(94) & 3 & 1 & 3 & A$_2$ & 200.0290242(47) \\
        44.739796(95) & 2.9105$\times10^{-3}$ & 3 & 2 & 1 & B$_1$ & 296.041254(67) & 3 & 1 & 2 & B$_2$ & 251.301459(67) \\
        168.1638(50) & 8.7250$\times10^{-2}$ & 4 & 2 & 3 & B$_1$ & 419.4653(50) & 3 & 1 & 2 & B$_2$ & 251.301459(67) \\
        115.73401(13) & 1.5816$\times10^{-2}$ & 4 & 1 & 3 & A$_2$ & 399.049356(97) & 3 & 2 & 2 & A$_1$ & 283.315346(94) \\
        240.0720(50) & 2.4471$\times10^{-1}$ & 4 & 3 & 1 & A$_2$ & 523.3873(50) & 3 & 2 & 2 & A$_1$ & 283.315346(94) \\
        21.2099(50) & 6.8544$\times10^{-6}$ & 4 & 1 & 4 & B$_2$ & 317.2512(50) & 3 & 2 & 1 & B$_1$ & 296.041254(67) \\
        \hline
    \end{tabular}
\end{table*}

\section{Conclusion}

We have updated three \textsc{marvel} networks for H$_3^+$, H$_2$D$^+$ and D$_2$H$^+$ to include all currently published spectroscopic data for these molecules.
We have also performed variational nuclear motion calculations using the program \textsc{DVR3D} for D$_2$H$^+$ and D$_3^+$ to produce the new MiZo line lists.
The empirical energy levels derived from the \textsc{marvel} networks have been used to update the calculated levels of their respective molecules.
This allows for the subset of transitions involving these \textsc{marvel}ised energies to be determined to much higher accuracy, making the use of these line lists well suited for high-resolution spectroscopy.
The D$_3^+$ calculations have been combined with a set of energies derived from effective Hamiltonian calculations computed using experimentally determined molecular constants.
Given these effective Hamiltonian constants were derived from transitions in infrared bands, the new D$_3^+$ line list is best suited for infrared studies.
Overall, the new D$_3^+$ line list is of lower resolution than the other three, due to the lack of \textsc{marvel}ised energy levels.

Hyperfine effects are known to be present in the spectra of H$_3^+$ and its deuterated isotopologues, due to the nuclear spin of the component protons and deuterons \citep{97JePaSp.H3+}.
These hyperfine splittings have not yet been observed however in either an astrophysical setting or in the laboratory.
Were experiments to be conducted to measure the hyperfine-resolved spectra of the species considered here, it would enable the construction of a hyperfine-resolved \textsc{marvel} network and subsequently line lists.
Similarly, any further measurements of hyperfine-unresolved spectra could be added to the current \textsc{marvel} network to further constrain the resultant empirical energy levels and hence improve the accuracy of the line lists.
If a sufficiently large number of spectra were observed for D$_3^+$ to form a well-connected spectroscopic network, it would enable us to further update the D$_3^+$ line list with \textsc{marvel} energies to a high-resolution standard.

The four line lists presented here, each consisting of a states file and transitions file, are made available via www.exomol.com.

\section*{Acknowledgements}
We thank Tibor Furtenbacher and Attila Cs\'asz\'ar for supplying the \textsc{marvel4} code and for helpful discussions during the course of this work.
This work was supported by the European Research Council (ERC) under the European Union’s Horizon 2020 research and innovation programme through Advance Grant number 883830 and the UK STFC under grant ST/R000476/1.
BP acknowledges support from the US National Science Foundation (CHE-1665370), and the Robert A. Welch Foundation (D-1523).
NFZ and OLP acknowledge support by State Project IAP RAS No. 0030-2021-0016.
JS is grateful to NKFIH for support (PD142580).

\section*{Data Availability}
The updated \textsc{MARVEL} input files and resulting output energy levels are given as supporting material.
All other data are available via the www.exomol.com website.

\bibliographystyle{mnras}

\label{lastpage}
\end{document}